\documentclass[%
 reprint,
 amsmath,amssymb,
 aps,
]{revtex4-2}

\usepackage{graphicx}
\usepackage{dcolumn}
\usepackage{bm}
\usepackage{xcolor}

\begin{document}

\preprint{APS/123-QED}

\title{A Universal Framework for Reconstructing Complex Networks and Node Dynamics from Discrete or Continuous Dynamics Data}

\author{Yan~Zhang}
\thanks{These authors have contributed equally to this work}
 \affiliation{School of Systems Science, Beijing Normal University, Beijing 100875, China}

\author{Yu~Guo}%
\thanks{These authors have contributed equally to this work}
\affiliation{%
 Software Institute, Nanjing University, Nanjing 210093, China
}%

\author{Zhang~Zhang}
\affiliation{
School of Systems Science, Beijing Normal University, Beijing 100875, China
}%
\author{Mengyuan~Chen}
\affiliation{%
China TravelSky Mobile Technology Co., Ltd, Beijing 100029, China
}%
\author{Shuo~Wang}
\affiliation{%
School of Systems Science, Beijing Normal University, Beijing 100875, China
}%
\author{Jiang~Zhang}
 \email{zhangjiang@bnu.edu.cn}
\affiliation{%
School of Systems Science, Beijing Normal University, Beijing 100875, China;\\ Swarma Research, Beijing, 102308, China
}%

\date{\today}

\begin{abstract}
Many dynamical processes of complex systems can be understood as the dynamics of a group of nodes interacting on a given network structure. However, finding such interaction structure and node dynamics from time series of node behaviours is tough. Conventional methods focus on either network structure inference task or dynamics reconstruction problem, very few of them can work well on both. This paper proposes a universal framework for reconstructing network structure and node dynamics at the same time from observed time-series data of nodes. We use a differentiable Bernoulli sampling process to generate a candidate network structure, and use neural networks to simulate the node dynamics based on the candidate network. We then adjust all the parameters with a stochastic gradient descent algorithm to maximize the likelihood function defined on the data. The experiments show that our model can recover various network structures and node dynamics at the same time with high accuracy. It can also work well on binary, discrete and continuous time-series data, and the reconstruction results are robust against noise and missing information. 

\end{abstract}

\maketitle


\section{Introduction}
{L}{iving} cells, brains, human society, stock markets, global climate systems, and so forth are complex systems composed of many nonlinear interactive units~\cite{boccaletti2006complex,watts1998collective,runge2019detecting}. By decomposing a complex system into a static network with dynamics on nodes, networked dynamical system models are powerful tools to describe complex systems, playing a paramount role in understanding their collective behaviours and controlling their functions~\cite{boccaletti2006complex,liu2016control}. However, building such models requires professional prior knowledge and modelling experience, which hinders the wide application of these methods. The reconstruction of such networked dynamical systems in a data-driven way remains a fundamental problem, i.e., to retrieve the interaction network structure and the node dynamics from time-series data of complex system behaviours without any subjective biases~\cite{wang2016data,ha2020deep}.

Historically, recovering network structure, and predicting node dynamical behaviors reconstruction or simulation for predictions for future states are different tasks that are studied separately. First, revealing network structure from time-series data is of great importance because the inferred network can not only help us to understand the behaviours of the system but also can provide information on inner causal structure of interactions within a complex system~\cite{runge2019detecting,pearl2009causality, tank2018neural,lowe2020amortized,glymour2019review}. The interdependence relations or causal structure can be obtained by directly calculating some statistical measures~\cite{peng2009partial}, perturbing the system~\cite{nitzan2017revealing}, optimising a score function~\cite{liu2018functional,runge2018causal}, or expanding the complex interaction dynamics on a set of basal functions~\cite{wang2016data,casadiego2017model,li2017reconstruction}, and other methods~\cite{granger1969investigating}. For example, ARNI (the algorithm for revealing network interactions) method can not only infer a network with high accuracy but also be adopted for various nonlinear dynamics. However, one disadvantage is that the performance of the model strongly depends on the choice of the basal functions. If the prior biases on basal functions are missing, this approach becomes very time-consuming, limiting its application to larger systems. 

Another branch of study is dynamics reconstruction from time series data. According to Taken's theorem, any dynamical behaviors of a non-linear system can be reconstructed from the time series data, once the observed sequence is long enough\cite{takens1981detecting}. Thereafter, various techniques have been developed to build models to simulate the dynamical behaviors behind data and to predict future states of the system~\cite{brockwell1991time,jaeger2001echo,schrauwen2007overview}. However, the techniques of conventional time series forecasting can hardly extend to complex systems because the task becomes very difficult as the number of variables to be predicted become very large and the long-range correlations between nodes widely exist. The problem can be alleviated when the interaction network structure between nodes are provided as inductive bias, graph (neural) network (GNN) models~\cite{scarselli2008graph,battaglia2018relational,wu2020comprehensive,sanchez2018graph,kipf2018neural,zhang2019neural,zheng2020gman} are designed particularly on this task. By learning complex functions of information aggregation and propagation on a
given network, GNNs can simulate any complex dynamics on such networks. However, a complete graph is always required for most GNN models, which hinders their wider applications~\cite{ha2020deep,kipf2018neural,alet2019neural,zhang2019general,pareja2020evolvegcn}.

Very few studies have been proposed to perform both network reconstruction and node dynamics reconstruction tasks together at the same time, although some network inference algorithms are capable of forecasting node states and the implicit and time-variant network structures can also be obtained from deep learning models for forecasting based on an attention mechanism~\cite{ha2020deep,pareja2020evolvegcn,velivckovic2017graph,wang2020learning,ijcai2019-264}. The first framework to derive an explicit network is NRI (neural relation inference)~\cite{kipf2018neural}, in which an encoder–decoder framework is used. However, the complicated encoding process to infer the connections from time series data has limited its scalability and accuracy on larger networks~\cite{zhang2019general,alet2019neural,ayed2019learning}.

This paper proposes a unified framework for automated interactions and dynamics discovery (AIDD). This is a universal framework for learning both the interaction structure and dynamics of a complex system from time series data. The design of a lightweight network generator and a universal dynamics learning component based on Markov dynamics makes AIDD not only be applicable to various networks and dynamics, but also enables it to reconstruct very large networks with high accuracy and robustness. The entire framework is differentiable so that it can be optimised directly by automatic differentiation and machine learning techniques~\cite{baydin2018autodiff}. Beyond tasks of network inference and time series forecasting, we propose a new method to test a learned data-driven model based on control experiments. Finally, we test the validity of our framework on real gene regulatory networks under noisy, incomplete, and interrupted data, which is close to realistic situations. The results demonstrate that high performance can be obtained.

\section{Methods}
\subsection{Problem Formulation}
Suppose the complex system to be considered evolves under discrete time steps. Thus, the dynamics to be reconstructed can be described by:
\begin{equation}
\label{eq.discrete}
    \mathbf{X}^{t+1}=f(\mathbf{X}^{t},A)+\mathbf{\zeta}^t,
\end{equation}
where $\mathbf{X}^t=[X_1^t,X_2^t,\cdot\cdot\cdot,X_N^t]\in \mathbb{R}^{N\times D}$ is the state of the system at time $t$, $N$ is the number of nodes, $D$ is the dimension of the state space of every single node, $A$ is the adjacency matrix of the interaction network to be reconstructed, and $\mathbf{\zeta}^t\in \mathbb{R}^{N\times D}$ is the noise imposed on nodes. However, Equation (\ref{eq.discrete}) can only describe the dynamical processes with explicit mathematical forms and cannot be applied to those defined by rule tables or transitional probabilities, such as cellular automata, Boolean dynamics, or Markov chains. Therefore, instead of Equation (\ref{eq.discrete}), we use a more general form, a Markov chain $\{\mathbf{X}^t\}$, to describe the dynamics.
\begin{equation}
\label{eq.markoviandynamics}
    f(\mathbf{X}^{t+1}=\mathbf{x}^{t+1}| \mathbf{x}^t,A)\equiv P(\mathbf{X}^{t+1}=\mathbf{x}^{t+1}|\mathbf{x}^t,A),
\end{equation}
where $f$ is the dynamics to be discovered, $\mathbf{x}^t = ({x_1}^t,{x_2}^t,\cdot\cdot\cdot,{x_n}^t)\in V^{N}$ is the observed time series, where $V$ is the state space of each single node and can be either $\mathcal{R}^D$ for continuous values or $\{0,1\}^D$ for discrete values. Here, we use $D$ dimensional one-hot vector to represent a finite set with $D$ elements. A one-hot vector is a vector with all 0s except one position taking 1, and if the position that 1 appears is $p$, then the vector represents the $p$-th element. E.g., $(0,0,1), (0,1,0)$ can represent $a,b$, respectively, which are the 1st and 2nd elements in set $\{a,b,c\}$. $P$ is the conditional probability. Equation (\ref{eq.markoviandynamics}) is compatible with \ref{eq.discrete} but more general~\cite{gardiner1985handbook}. It can even be extended to non-Markov random processes with finite histories by adding more hidden auxiliary variable variables~\cite{hochreiter1997long}.

However, it is difficult to infer the probabilities in
(\ref{eq.markoviandynamics}), particularly when $N$ is large. Fortunately, the interactions of complex systems are always localised, which means that $P(\mathbf{X}^{t+1}=\mathbf{x}^{t+1}|\mathbf{x}^t,A)$ can be factorised into local transitional probabilities~\cite{scholkopf2019causality}:
\begin{equation}
\label{eq.causalmodule}
    P(\mathbf{X}^{t+1}=\mathbf{x}^{t+1}|\mathbf{x}^t,A)=\prod \limits_{i=1}^N P(X^{t+1}_i = x^{t+1}_i|\mathbf{x}^t\odot A_{\cdot i}),
\end{equation}
where $\odot$ is the element-wise product, and $A_{\cdot i}$ represents the $i^{th}$ column of matrix $A$, and $\mathbf{x^t}\odot A_{\cdot i}$ is a vector representing the state combination of all neighbour nodes of $i$. Then
\begin{equation}
    f_i(X^{t+1}_i= x^{t+1}_i| \mathbf{x^t}\odot A_{\cdot i})\equiv P(X^{t+1}_i= x^{t+1}_i|\mathbf{x^t}\odot A_{\cdot i}) 
\end{equation}
represents the local dynamics of node $i$, which is also called a causal mechanism in the literature~\cite{scholkopf2019causality}. Where $\odot A_{\cdot i}$ can be regarded as a filter on all nodes' states $\mathbf{x}^t$, so $\mathbf{x}^t\odot A_{\cdot i}$ is the combination of the states of $i$'s neighbors. \\
To be notice, for most network dynamics, nodes always share a same node dynamic, that is, $f_i\equiv f$ for any $i$. Therefore, the dependence of $f$ on $\mathbf{x}^t\odot A_{\cdot i}$ need be a set function. That is, for any node, $f$ is permutation invariant on all its neighbors such that $f$ can be applied on neighborhoods with different number of nodes~\cite{zaheer2017deep,xu2018powerful}.

Therefore, our task becomes the reconstruction of the network $A$ and learning the local dynamics $f_i$ according to the observed time series $\mathbf{x}=(\mathbf{x}^1, \mathbf{x}^2,\cdot\cdot\cdot,\mathbf{x}^T)$ with $T$ time steps on multiple samples. 

\subsection{Model}
To finish the task, we build a parameterized model with neural networks to represent the reconstructed network structure and node dynamics.

\begin{figure*}[ht!]
\centering
\includegraphics[scale=0.7]{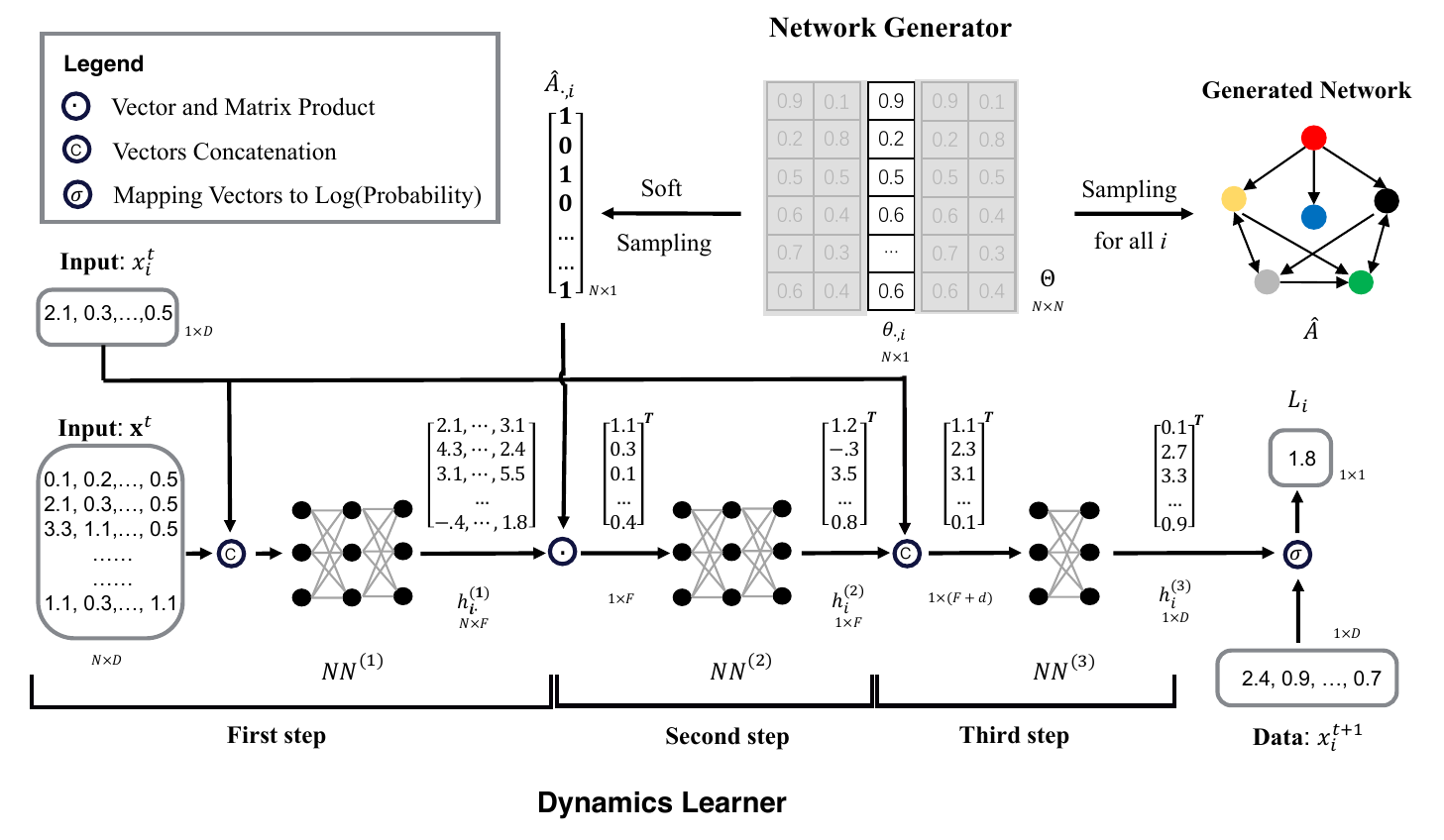}
\caption{
The architecture of the framework Automated Interaction and Dynamics Discovery(AIDD). The upper row outlines how the network generator samples a column of the adjacency matrix for node $i$, and reconstructs the entire network according to a $N\times N$ dimensional parameter matrix $\Theta$ with $i$-th $N$ rows column vector representing the probabilities of $\hat{A}_{\cdot i}$ taking value $1$ for node $i$; and the bottom row shows how the dynamics learner predicts the log-probability of node $i$'s state at the next time step. We also show all the input and output vectors(matrices) and their sizes beside each unit. The first step of the dynamics learner converts the input data of all nodes into their node representations $\mathbf{h}^{(1)}$, the second step multiplies the $i$-th column of the adjacency matrix $\hat{A}_{\cdot i}$ and the representations of all nodes $\mathbf{h}^{(1)}$ to feed into the second neural network $NN^{(2)}$ to generate a new representation $h_i^{(2)}$ with aggregated information of all neighbors of $i$; and the third step will concatenate $h_i^{(2)}$ and $x_i^t$ to feed into the third neural network $NN^{(3)}$ to generate $h_i^{(3)}$ which represents the predicted state of $i$ at the next step; after that, the log-likelihood will be calculated according to $h_i^{(3)}$ and $x_i^{t+1}$. Finally, the back-propagation algorithm will be automatically executed by the deep-learning framework(e.g., PyTorch). Each neural network $NN^{(1,2)}$ is a three layered feed-forward network with $F=256$ hidden units and ReLU activate function, while $NN^{(3)}$ has two layers.}
\label{fig:framework} 
\end{figure*}

Our model consists of two parts, as shown in Figure \ref{fig:framework}. The first part is a network generator that can output a candidate network adjacency matrix $\hat{A}$ according to the parameters $\Theta=\{\theta_{ij}\}$. The second part then attempts to simulate the node dynamics $f_i$ by using a set of neural networks $\hat{f}_{\phi_i}$ for any node $i$, which is parameterized by $\Phi=(\phi_1,\cdot\cdot\cdot\phi_N)$ to predict the future state $\hat{X}_i^{t+1}$ according to the candidate matrix $\hat{A}$ and the observed state of the previous time step $\mathbf{x}^t$. 

\subsubsection{Network Generation}
Instead of using the complicated neural network architecture to generate the candidate network as described in~\cite{kipf2018neural}, we directly sample each element in the adjacency matrix $\hat{A}_{ij}\sim Bernoulli(\theta_{ij})$, where $\theta_{ij}\in [0,1]$ represents the probability that the entry of the $i^{th}$ row and the $j^{th}$ column in $\hat{A}$ takes the value $1$. To make the sampling process differentiable, we use the Gumbel-softmax technique~\cite{jang2016categorical,kipf2018neural} to generate the adjacency matrix~\cite{zhang2019general}.
\begin{equation}
\label{eq.gumbel}
    \hat{A}_{ij}=\frac{g(\theta_{ij}, \xi_{ij},\tau)}{g(\theta_{ij}, \xi_{ij},\tau)+g(1-\theta_{ij}, \xi'_{ij},\tau)},
\end{equation}
where
\begin{equation}
\label{eq.Z}
    g(\theta, \xi, \tau)=\exp\left[\frac{\log\left(\theta\right)+\xi}{\tau}\right],
\end{equation}
and $\xi_{ij},\xi'_{ij}\sim Gumbel(0,1)$ are random numbers following the standard Gumbel distribution, and $\tau$ is the parameter of temperature to adjust the evenness of distribution of the sampling results which is always $1$ in this paper. The random numbers generated by (\ref{eq.gumbel}) have a similar distribution as $Bernoulli(\theta_{ij})$, especially when $\tau$ is small. The simulated sampling process is differentiable such that the gradients can be passed by. When $\tau\rightarrow 0$, $\hat{A}_{ij}$ exactly equals 1 with probability $\theta_{ij}$, or 0 with probability $1-\theta_{ij}$. 

In this way, our framework has a large improvement in flexibility and computational efficiency compared to the encoder-decoder frameworks such as ~\cite{kipf2018neural}
and can be applied to very large networks. Further, the introduction of noise $\xi_{ij}$ can push the network generator to jump out of local minimums during optimisation.

\subsubsection{Node Dynamics Simulation}

According to (\ref{eq.causalmodule}), the dynamics learning part can also be decomposed into local dynamics node by node. Each local node dynamics can be modeled by a feed-forward neural network.

To simulate very complex non-linear node dynamics in various types for each node, we adopt graph neural network framework. Graph neural network(GNN) is a special kind of neural networks that act on graph-structured data\cite{scarselli2008graph}. It constructs a parameterized message passing process on a given graph structure, compares it with the real network dynamic process, and adjust the parameters by the back-propagation algorithm to simulate the real network dynamics process\cite{kipf2018neural}.

In our GNN framework, we decompose the message passing process into three steps as shown in Figure \ref{fig:framework}. For any node $i$, the first step is to encode its interaction with any other node $j\in [0,N]$ to form an abstract representation vector:
\begin{equation}
    \label{eq:edgeencoding}
    h_{ij}^{(1)}=NN^{(1)}(x_i^t,x_j^t)
\end{equation}
with dimension $F$(in this paper, we set $F=128$). By iterating this computation for all $j$, we obtain a matrix:
\begin{equation}
    \label{eq:aggregation}
    h_{i\cdot}^{(1)}=(h_1^{(1)},h_2^{(1)},\cdot\cdot\cdot, h_N^{(1)})
\end{equation}
with dimension $F\times N$. This hidden variable represents all possible interactions with $i$.

The second step is to filter out the irrelevant information of non-adjacent nodes of node $i$ according to the candidate adjacency matrix generated in the previous step $\hat{A}_{.,i}$:

\begin{equation}
    \label{eq:projection}
    h_i^{(2)}=NN^{(2)}\left( h_{i\cdot}^{(1)}\cdot \hat{A}_{.,i}\right)
\end{equation}
where $h_i^{(2)}$ with dimension $F$ is also the abstract representation vector of $i$ after aggregating the information of all $i$'s neighbors. The last step is to generate a new hidden representation vector prepared for output:
\begin{equation}
    \label{eq:output}
    h_i^{(3)}=NN^{(3)}\left(x_i^t, h_i^{(2)}\right).
\end{equation}

Here, we also feed $x_i^t$ into $NN^{(3)}$ to promote prediction accuracy. In equations \ref{eq:aggregation}, \ref{eq:projection}, and \ref{eq:output}, $NN^{(k)}$ for $k=1,2,3$ are feed-forward neural networks with parameters $\{\phi^{(k)}\}$ in the nearly identical structures (with $F$ hidden units and ReLU activate function. $NN^{(1,2)}$ have three layers, $NN^{(3)}$ has two layers). Notice that, the parameters $\phi^{(k)}$ are shared for all nodes ($\phi_i^{(k)}=\phi^{(k)}$ for any node $i$ and step $k$) which can reduce the number of parameters in a large sense. As shown in ~\cite{zaheer2017deep,xu2018powerful}, Equations \ref{eq:aggregation}, \ref{eq:projection}, and \ref{eq:output} can simulate any network dynamics with the form of Equation \ref{eq.causalmodule}.

Finally, we will map the hidden representation of $i$: ${h}_{i}^{(3)}$ into the logarithm of the probability (log-likelihood) that the node state $\hat{X}_i^{t+1}$ taking the value $x$:
\begin{equation}
    \label{eq:probability}
    \sigma\left(x;h_i^{(3)}\right)\equiv \log\hat{f}_i\equiv \log P\left(\hat{X}_i^{t+1}=x|\mathbf{x}^t\odot \hat{A}_{\cdot i}\right),
\end{equation}
where, $\sigma$ is the function to map ${h}_{i}^{(3)}$ and $x_i$ to the log-probability, and the functional form depends on the value range $V$ of node state. 

If $V=\mathbb{R}^D$ where $D$ is the dimension of values, then $\sigma$ has the form of logarithm of the standard normal distribution $\log \mathcal{N}(\mathbf{\mu},\mathbf{1})$ with the mean vector $\mathbf{\mu}=h_i^{(3)}$ as the output of $NN^{(3)}$, and the variance as the value one (or we can also consider log-laplacian form $\log \mathcal{L}(\mathbf{\mu},\mathbf{1})$, and different distributions will affect the form of objective function as discussed in the next sub-section). 

If $V=\{0,1\}^D$, $h_i^{(3)}$ is a one-hot vector representing for discrete values, then $\sigma$ is a log-softmax function: $\sigma(x;{h}_{i}^{(3)})=\log\left[\exp(h_i^{(3)})/\sum_{j=1}^D\exp(h_j^{(3)})\right]$ and the output $h_i^{(3)}$ of $NN^{(3)}$ represents the logarithms of the probabilities that $\hat{X}_i^{t+1}$ taking each possible value. 

To generate node states at time step $t+1$, we can sample values according to the probability $P(\hat{X}_i)$ for node $i$ independently. In this way, we can simulate one step dynamics of all nodes.

In general, the first step is to encode input information of each node and possible neighbors into representations, the second step is to aggregate the information from neighbors, and the last step is to generate output. This design has been verified to be suitable for learning various complex network dynamics~\cite{zhang2019general}. 

\subsection{Objective function}
With the model which can generate candidate network and node dynamics, we can convert the network and node dynamics reconstruction problem as an optimization problem by minimising the following objective function.

\begin{equation}
\label{eq.object}
\begin{aligned}
   & L(\Theta, \Phi)=\\
   & \mathbb{E}_{\hat{A}\sim B(\Theta)}\left(-\sum\limits_{t=1}^T \log P(\hat{\mathbf{X}}^{t+1}=\mathbf{x}^{t+1}|\mathbf{x}^t,\hat{A},\Phi)\right)+\lambda \sum_{ij}\hat{A}_{ij}\\
   & \approx -\frac{1}{K}\sum_{k=1}^{K}\sum_{t=1}^{T-1}\sum_{i=1}^{N} L_i+\lambda \sum_{ij}\hat{A}_{ij}&,
\end{aligned}
\end{equation}
where
\begin{equation}
\label{eq:localobj}
    L_i=\log\hat{f}_i(x_{i}^{t+1})\equiv \log P\left(\hat{X}_i^{t+1}=x_{i}^{t+1}|\mathbf{x}^t\odot \hat{A}_{\cdot i}^k; \phi\right)
\end{equation}
is the local log-likelihood, and $K$ is the number of samples for matrix $\hat{A}$ under a given $\Theta$, $\hat{A}^k$ is the $k$-th sample of adjacency matrix, and $x_{i}^t$ is the observational vector of state of node $i$ at time $t$. $B$ represents Bernoulli distribution. We sample $K$ candidate networks under the same parameters $\Theta$ to evaluate dynamics simulator to improve accuracy. Thus, the objective function contains two terms, the former being the log-likelihood, which can be decomposed into local log-likelihood. The latter is the structural loss to conform the network to be sparse while avoiding over-fitting. The parameter $\lambda$ can adjust the relative importance of the structural loss. 

It can be easily derived that if the state space $V=\mathbb{R}^D$, and $\sigma$ takes form of Normal or Laplacian distribution, then the local log-likelihood (\ref{eq:localobj}) is:
\begin{equation}
\label{eq:nodeobj}
    L_i= \Vert x_{i}^{t+1}-h_i^{(3)} \Vert_p,
\end{equation}
where $p=2$ if Gaussian distribution is adopted and $L_i$ is mean-square error(MSE), and $p=1$ if Laplacian distribution is used and $L_i$ is mean-absolute error(MAE). And $h_i^{(3)}$ is the output of the last layer which represents the mean vector $\mathbf{\mu}$ of the Gaussian(Laplacian) distribution. 

Then, the network dynamics to fit the observational data can be obtained by tuning the parameters of $\theta_{\cdot i}$ in candidate adjacency matrix and $\phi$ in dynamics simulators to optimise the objective functions (\ref{eq:nodeobj}) node by node~\cite{lee2018deep}. We use the stochastic gradient descent algorithm to optimise. 

\subsection{Training}
To train the model, we separate the time series data into data pairs with the following format $(\mathbf{x}^t,\mathbf{x}^{t+1})$ for any time step $t$. We feed $\mathbf{x}^t$ into the neural network model of any node $i$ (selected one by one) to get the prediction of $i$'s node state at next time $\hat{x}_i^{t+1}$, and this will be compared with the target $x_i^{t+1}$. The objective function will be evaluate and to train all the parameters. The computation can be parallel in batches, we group data pairs $(\mathbf{x}^t,\mathbf{x}^{t+1})$ into batches. In the experiments, we sample the candidate network at each train epoch, that is equivalent to set $K$ as the total number of training steps. After forward computation, the parameters in the model will be updated according to stochastic gradient decent algorithm. We do not need to derive the explicit expression of the gradients, the platforms like PyTorch or Tensorflow can calculate the gradients automatically by using automatic differentiation technique. 
See Supplemental Material at section 1 for more details about training algorithm.

\subsection{Performance Metrics}
We separate all the time series data into training, validation, and test for training the model, selecting the hyper-parameters, and testing the performance of the model, respectively. 
See Supplemental Material at section 4 for the ratios of these three data sets are various for dynamics, which includes Refs.~\cite{kipf2018neural, garcia2002coupled,garcia2002coupled,campbell1954voter,karlebach2008modelling,brockmann2013hidden,schaffter2011genenetweaver}.

In order to evaluate the effectiveness of our model in network inference problem and dynamics prediction task, we use the following indicators on the test set.

\textbf{AUC: (Area Under Curve)} It evaluates the accuracy of the reconstructed network. It is defined as the area under the ROC curve and is an index for comprehensive evaluation of accuracy and recall. We also use AUC to evaluate the performance of the dynamics reconstruction for binary dynamics.

\textbf{MSE: (Mean Squared Error)} It measures the performance of the node dynamics reconstruction. It is the expectation of the square of the difference between the true value and the predicted value (estimated value).

\textbf{MAE: (Mean Absolute Error)} It measures the performance of the node dynamics reconstruction. It is the expectation of the absolute of the difference between the true value and the predicted value (estimated value).

\section{Results}
\subsection{Experimental Design}
To test our method on various network structures and dynamics, we use different well-known dynamics on both real and modeled networks to generate time series data.
\subsubsection{Networks}
\begin{table}[h!]
\caption{Network Parameter of Real Network. For each network, $N, E, \langle k\rangle$ and $\langle c\rangle$ represent the number of nodes, the number of edges, average degree, and average clustering coefficient, respectively.\\
}
\centering
\begin{ruledtabular}
\begin{tabular}{cccccc}
   {Network} & {$N$ }& {$E$} &$\langle k\rangle$&$\langle c\rangle$&Directed\\
  \hline
  {Email} &1,133 &10,902 & 4 & 0.220&no  \\
  {Road} &1,174 &2,834 & 2 & 0.078&no  \\
  {Blog} &1,224 &19,025 & 15 & 0.210&yes \\
  {Dorm} &217 &2,672 & 24 & 0.399&yes \\
  {City} & 371 & 2,752& 7& 0.485 & yes \\
  {Gene} & 100 & 195& 2& 0.070& yes\\ 
\end{tabular}
\end{ruledtabular}
\label{tab.realnetwork}
\end{table}
The networks we used are shown in Table \ref{tab.realnetwork}. These graphs are either generated by well known models (ER~\cite{erdHos1960evolution} for Erdos Renyi, WS~\cite{watts1998collective} for Watts-Strogatz, and BA~\cite{barabasi1999emergence} for Barabasi-Albert) or from empirical data including a gene network (for S. cerevisiae, Gene)~\cite{schaffter2011genenetweaver}, an inter-city traffic network (City) of China, three social networks (Email, Dorm, and Blog), and a road network(Road)~\cite{nr}.
See Supplemental Material at section 3 for more details for empirical networks, which includes Refs.~\cite{nr,zhang2020investigating,schaffter2011genenetweaver}.
In the experiments where the number of nodes is less than 1000, the weight $\lambda$ (see Equation \ref{eq.object}) is set to 0.0001, while $\lambda$ is set to 0.001 for larger networks. 
\subsubsection{Test Dynamics}
We generate time series data by a set of network dynamics including continuous, discrete, and binary dynamics. As shown in Table \ref{tab:table_dyn}, Spring (spring dynamics)~\cite{kipf2018neural}, SIR (an inter-city meta-population SIR epidemic model)~\cite{brockmann2013hidden}, and Michaelis–Menten kinetics~\cite{karlebach2008modelling} all are examples of continuous dynamics. The coupled map network (CMN)~\cite{garcia2002coupled} and the voter model~\cite{campbell1954voter} are representatives of discrete and binary dynamics, respectively. The equations for describing these dynamics are reported in Table \ref{tab:table_dyn}, and more details are referred to Supplementary Material section 4. Notice that all the dynamics listed in the table can be converted into a Markovian dynamics.
\begin{table*}
\caption{Functions for various dynamical rules, where $X_i$ is the state of $i$, $N_i$ represents the neighbors of $i$.}
\begin{ruledtabular}
\begin{tabular}{cccc}
Type & Model & Dynamics & Description\\
\hline
Continuous & Spring & {$\frac{d\mathbf{r}_i}{dt}=\mathbf{v}_i$} &{where, $X_i\equiv \{\mathbf{r}_i,\mathbf{v}_i\}$, $\mathbf{r_i,v_i}\in \mathbb{R}^2$}\\
{} & {} & {$\frac{d\mathbf{v}_i}{dt}=-k\sum_{j \in N_i}(\mathbf{r}_i-\mathbf{r}_j)$}&{$k=0.1$} \\
\hline
{Continuous} & Michaelis–Menten kinetics &  {$ \frac{dX_i}{dt}=-X_i+\frac{1}{|N_i|}\sum_{j\in N_i} \frac{X_j}{1+X_j}+\xi _i$}&{$X_i\in \mathbb{R}$, $\xi_i\sim \mathcal{N}(0,0.01)$}\\
\hline
{Continuous} & SIR & {$    \frac{ds_n}{dt}=-as_n i_n+\omega\sum\limits_{m\neq n}P_{nm}(s_m-s_n)$}&{where, $X_n\equiv \{s_n,i_n,r_n\}\in [0,1]^3$}\\
{} & {} & {$ \frac{di_n}{dt}=as_n i_n-bi_n+\omega\sum\limits_{m\neq n}P_{nm}(i_m-i_n)$ }&{$a=0.27,b=0.12,\omega=0.02$}\\
{} & {} & {$\frac{dr_n}{dt}=bi_n+\omega\sum\limits_{m\neq n}P_{nm}(r_m-r_n)$}&{$P_{nm} \in [0,1]$(see section 4 of SI)}\\
\hline
Discrete & Coupled Map Network & {$ X_{i}^{t+1}=(1-\epsilon)f(X_i^t)+\frac{\epsilon}{|N_i|}\sum\limits_{j=1}^{N} f(X_{j}^t)$}&{where, $X_i\in [0,1]$}\\
{} & {} &{}&{$f(X_i^t)=\eta X_i^t(1-X_i^t)$, $\eta = 3.5, \epsilon = 0.2$}\\  
\hline
Binary & Voter & {$P\left(X_i^{t+1}=1|{X^t_{j \in N_i}}\right)=\frac{\sum_{j \in N_i}X_j^{t}}{|N_i|}$}&{$X_j \in \{0, 1\}$}\\

\end{tabular}
\end{ruledtabular}
\label{tab:table_dyn}
\end{table*}

\subsection{Experimental Results}
To generate time series data, we ran the simulation for several time steps for each dynamical model and network. All the reported results are on the testing dataset. 
We compare our model to a series of baseline methods for both network inference and single step forecasting tasks. ARNI~\cite{casadiego2017model} and NRI~\cite{kipf2018neural} are both the state-of-the-art models for network inference and time series forecasting. The former is based on the block-orthogonal regression method, and the latter is based on deep learning and graph networks. Two other frequently used statistical metrics, partial correlation and mutual information, are also compared on the network inference task. In addition, the classic time series forecasting model, long short-term memory (LSTM)~\cite{hochreiter1997long}, is compared on the prediction task. 
See Supplemental Material at section 2 for the details of compared methods, which inculdes Refs.~\cite{casadiego2017model,kipf2018neural,hochreiter1997long}.
The comparison results on various networks and dynamics are shown in Table \ref{table_com}.

\begin{figure}[ht!]
\centering
\includegraphics[scale=0.54]{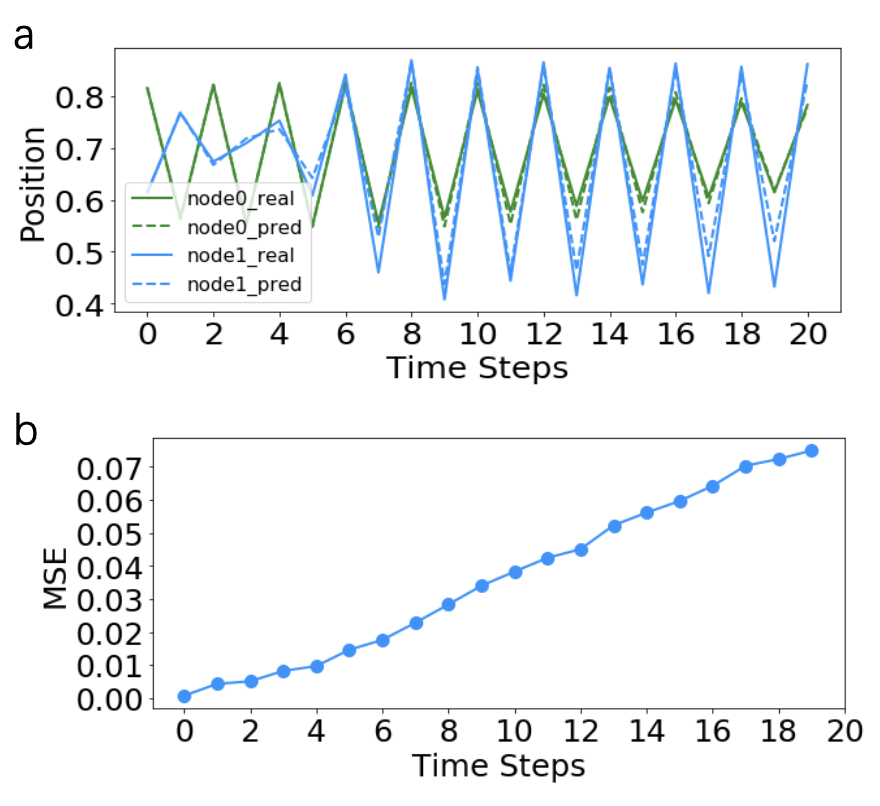}
\caption{Multi-step prediction results of AIDD on CMN dynamics data in a 10-node ER network. In (a), we show the time series data of multi-step predictions and the ground truths for two selected nodes. In (b), we show how the mean square error (MSE) increases with time for CMN dynamics. The parameters are the same as in Table \ref{table_com}. }
\label{fig:dyn_learn}
\end{figure}

\begin{table*}[]
\caption{Comparisons of performance on network inference and dynamics prediction tasks between AIDD and other selected methods (columns) on different dynamics and networks (rows). In the network column, we use network - size format. The networks marked with ``D'' represent directed graphs. The same data volume is shared for different methods in one row. The items marked by ``-'' indicate that valid results of the model cannot be obtained due to the limitations of the specific method on dynamics, memory, or time consumption. The best results among all the compared algorithms in the same row are boldfaced, and the second-best results are marked ``*''. All networks generated by models share the same edge density value, which is 1\% for large networks (size $>$ 10), and it is 20\% and 3\% for small networks with sizes smaller than 10, and ER networks with size = 200, respectively, to avoid isolated nodes. All the results are the averages of five repeated experiments. }
\setlength{\tabcolsep}{0.05mm}{
\begin{ruledtabular}
\begin{tabular}{cccccccccccc}

\multicolumn{1}{l}{Type} & Model & Network & \multicolumn{2}{c}{ARNI} & MI & PC & \multicolumn{2}{c}{NRI} & LSTM & \multicolumn{2}{c}{OURS} \\ 
\multicolumn{1}{l}{} &  &  & AUC & MSE & AUC & AUC & MSE/AUC\_Dyn & AUC & MSE/AUC\_Dyn & MSE/AUC\_Dyn & AUC \\ 
\colrule
Continuous & Spring & ER-10 & 0.5853 & $1.33 \times 10^{-3}$ & 0.7500 & 0.8250 & $\pmb{2.60 \times 10^{-8}}$ & 0.9998* & $2.98 \times 10^{-4}$ & $2.70 \times 10^{-4}$* & \textbf{1.0} \\ 
 &  & WS-10 & 0.5125 & $1.58 \times 10^{-3}$ & 0.6875 & 0.7875 & $\pmb{8.40 \times 10^{-8}}$ & 0.9997* & $3.35 \times 10^{-4}$ & $3.31 \times 10^{-4}$* & \textbf{1.0} \\ 
 &  & BA-10 & 0.5169 & $1.10 \times 10^{-3}$ & 0.6422 & 0.6571 & $\pmb{7.00 \times 10^{-10}}$ & 0.9999* & $2.14 \times 10^{-4}$ & $2.90 \times 10^{-5}$* & \textbf{1.0} \\ 
 &  & ER-2000 & - & - & 0.4997 & - & - & - & $2.25 \times 10^{-3}$& $\pmb{1.18 \times 10^{-5}}$ & \textbf{0.9886} \\ 
 &  & WS-2000 & - & - & 0.5002 & - & - & - & $5.89 \times 10^{-3}$ & $\pmb{8.51 \times 10^{-6}}$ & \textbf{0.9933} \\ 
 &  & BA-2000 & - & - & 0.5010 & - & - & - & $4.54 \times 10^{-3}$ & $\pmb{2.09 \times 10^{-3}}$ & \textbf{0.9523} \\ 
 & SIR & City-371(D) & 0.5424* & $8.25 \times 10^{-3}$& 0.5027 & 0.5119 & - & - & $2.28 \times 10^{-3}$* & $\pmb{2.98 \times 10^{-5}}$ & \textbf{0.9156} \\ 
 & Menten & Gene-100(D) & \textbf{1.0} & $9.71 \times 10^{-3}$ & 0.5416 & 0.6574 & - & - & $2.29 \times 10^{-3}$* & $\pmb{4.37 \times 10^{-5}}$ & 0.9960* \\ \hline
Discrete & CMN & ER-10 & \textbf{1.0} & $\pmb{2.33 \times 10^{-9}}$ & 0.5745 & 0.7804 & $1.40 \times 10^{-5}$ & 0.8850 & $2.60 \times 10^{-4}$ & $5.60 \times 10^{-6}$* & \textbf{1.0} \\ 
 &  & WS-10 & \textbf{1.0} & $\pmb{2.35 \times 10^{-9}}$ & 0.6875 & 0.8375 & $9.40 \times 10^{-6}$ & 0.9331 & $2.40 \times 10^{-4}$ & $2.80 \times 10^{-6}$* & \textbf{1.0} \\ 
 &  & BA-10 & \textbf{1.0} & $\pmb{2.40 \times 10^{-9}}$ & 0.4390 & 0.7439 & $1.30 \times 10^{-5}$ & 0.6753 & $9.21 \times 10^{-5}$ & $6.90 \times 10^{-6}$* & \textbf{1.0} \\ 
 &  & ER-200 & 0.8441* & $4.17 \times 10^{-2}$ & 0.5774 & 0.7648 & - & - & $5.91 \times 10^{-5}$* & $\pmb{2.04 \times 10^{-6}}$ & \textbf{0.9987} \\ 
 &  & WS-200 & \textbf{1.0} & $\pmb{2.36 \times 10^{-9}}$ & 0.6969 & 0.7506 & - & - & $1.63 \times 10^{-4}$ & $1.95 \times 10^{-6}$* & 0.9987* \\
 &  & BA-200 & 0.8840* & $2.45 \times 10^{-2}$ & 0.5533 & 0.7493 & - & - & $1.46 \times 10^{-4}$* & $\pmb{2.57 \times 10^{-6}}$ & \textbf{0.9874} \\ 
 &  & WS-1000 & - & \textbf{-} & 0.5670 & - & - & - & $3.54 \times 10^{-5}$ & $\pmb{2.92 \times 10^{-6}}$ & \textbf{0.9795} \\ 
 &  & BA-1000 & - & - & 0.5290 & - & - & - & $\pmb{3.46 \times 10^{-5}}$ & $5.48 \times 10^{-5}$ & \textbf{0.9105} \\ \hline
Binary & Voter & ER-10 & - & - & 0.4390 & 0.4552 & 0.9156* & 0.5305* & 0.5413 & \textbf{0.9664} & \textbf{1.0} \\
 &  & WS-10 & - & - & 0.4375 & 0.5250* & 0.8301* & 0.5196 & 0.6966 & \textbf{0.9452} & \textbf{1.0} \\ 
 &  & BA-10 & - & - & 0.4390 & 0.4607 & 0.9510* & 0.5192* & 0.5426 & \textbf{0.9853} & \textbf{1.0} \\ 
 &  & WS-1000 & - & - & 0.5470 & - & - & - & 0.5301 & \textbf{0.6567} & \textbf{0.9984} \\ 
 &  & BA-1000 & - & - & 0.5030 & - & - & - & 0.5236 & \textbf{0.6746} & \textbf{0.9573} \\ 
 &  & Email-1133 & - & - & 0.4999 & - & - & - & 0.5329 & \textbf{0.7286} & \textbf{0.9737} \\ 
 &  & Road-1174 & - & - & 0.5004 & - & - & - & 0.5448 & \textbf{0.8829} & \textbf{0.9999} \\ 
 &  & Dorm-217(D) & - & - & 0.5219 & - & - & - & 0.5688 & \textbf{0.6918} & \textbf{0.9920} \\ 
 &  & Blog-1224(D) & - & - & 0.4995 & - & - & - & 0.5300 & \textbf{0.6366} & \textbf{0.8401} \\ 
\end{tabular}
\end{ruledtabular}}
\label{table_com}

\end{table*}
The model can also work well on dynamics simulation. As shown in Table \ref{table_com}, the MSE errors for continuous and discrete time series are always smaller than $10^{-3}$ for single step prediction. The AUC accuracy for binary state dynamics can also get high values. 
The model can also output multi-step prediction results by feeding the result of the one-step prediction output back to the model as the input. Figure \ref{fig:dyn_learn} shows the results for the selected dynamics.

\subsection{Performance under Different Network Parameters}
\begin{figure*}[ht!]
\centering
\includegraphics[scale=0.5]{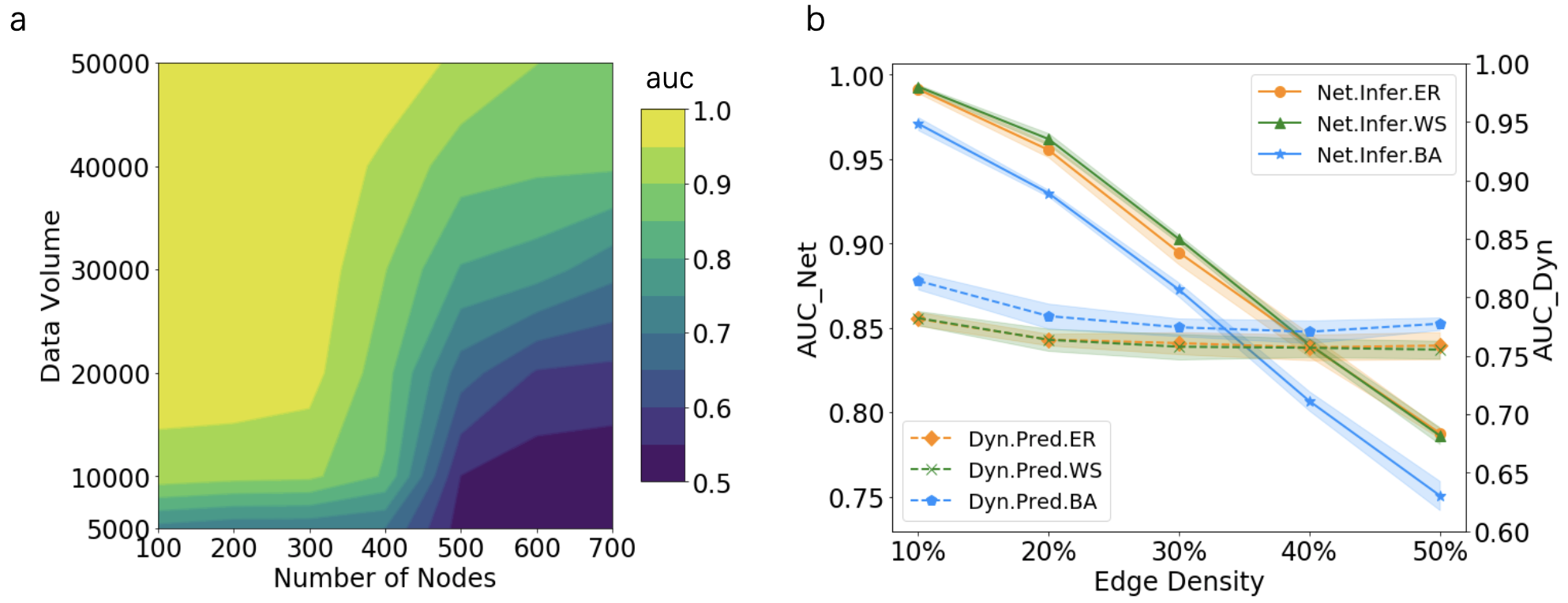}
\caption{The performance of AIDD under different factors in network inference and dynamics learning. (a) shows how the number of nodes and the volume of data (the number of samples$\times$ the number of time steps, which was fixed to 100) jointly influence the network inference accuracy on WS networks under CMN dynamics. With the exception of 100 nodes with a 4\% edge density, all nodes shared the same edge density value, which was 1\%.
(b) shows how performance decreases with edge density. For the experiments in (b), we set the number of nodes to 100, and the sparse matrix parameter $\lambda$ was set to 0.}
\label{fig:node_and_data}
\end{figure*}

In general, our model works very well on the large sparse networks, and the performance on both tasks decreases as the edge density increases, as shown in Fig. \ref{fig:node_and_data}(b).  

We can improve the accuracy by increasing the amount of data. Figure \ref{fig:node_and_data}(a) shows how the area under the curve (AUC) depends on both network size and data volume to be fed into the model systematically.
There is a trade-off between network size and data volume under a given accuracy, as shown in Fig. \ref{fig:node_and_data}(a). It is interesting to observe that data volume is sensitive to network size only when the number of nodes is between 300 and 500, and beyond that, a minimum amount of data volume is sufficient to obtain an acceptable accuracy (e.g., $AUC=0.7$), and this almost does not depend on how large the network is. We suppose that the given total number of epochs is sufficient for training the model only for networks smaller than 300, which is the reason why a sudden increase in data volume is observed. When the size is larger than 500, the model can converge quickly when sufficient data volume is provided; therefore, the curves become insensitive again.

\subsection{More Experiments}
To further test our method on different conditions, we design more experiments including network and dynamics reconstruction on incomplete networks, under the real scenario, and control experiment with learned network and node dynamics.
\subsubsection{Robustness against noise and hidden nodes}
\begin{figure*}[ht!]
\centering
\includegraphics[scale=0.66]{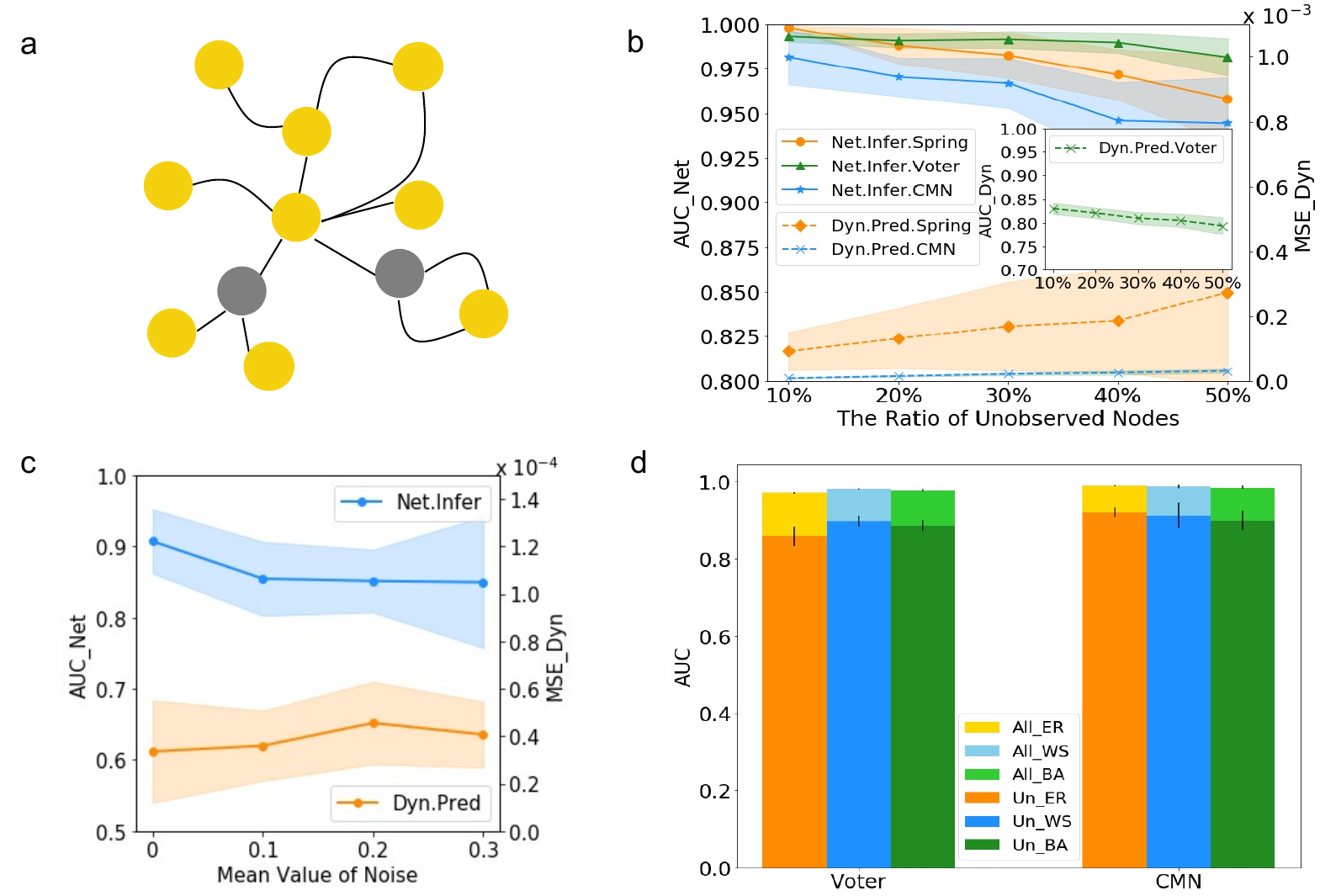}
\caption{The robustness evaluation of our model against noise and missing nodes. (a) shows a schematic ground truth network with missing information on the unobserved nodes (grey nodes). (b) shows the influence of proportion of unobserved nodes on the accuracy of interaction inference on the partial network with observed nodes measured by AUC, and the accuracy of dynamic predictions (inset) measured by the MSE of the observable nodes on Spring, CMN, and the AUC of the observable nodes on Voter dynamics. All the experiments were conducted on an ER network with 100 nodes, and all networks generated by models share the same edge density value, which is 4\%.  (c) shows the dependence of AUC and MSE on the mean of noise added on each node for the Michaelis–Menten kinetics (Gene dynamics) on the yeast S. cerevisiae gene network with 100 nodes. (d) shows the ability to infer interactions on the entire network (the light colour bars) and the unobserved partial networks (the dark colour bars). All the experiments are conducted on CMN and Voter dynamics with ER, WS, and BA networks, and all networks contain 100 nodes with 10\% unobservable nodes selected randomly, and all networks generated by models share the same edge density value, which is 4\%.}
\label{fig:robust}
\end{figure*}

A good data-driven model must be robust against noise and unobservable nodes such that it can be applied to the real world. To show the robustness against noise of AIDD, we plot changes in AUC with the magnitude of noise on Michaelis–Menten kinetics~\cite{karlebach2008modelling}, which can describe the dynamics of Gene regulatory networks, as shown in Fig. \ref{fig:robust}. Our model can recover the network structure with 0.85 AUC when the mean magnitude of noise is 0.3.

In real applications, we can only obtain partial information of the entire system owing to the limitations of observation. Thus, a certain proportion of nodes are unobservable or hidden. This requires the inference algorithm to be robust to hidden nodes. Thus, we test the AIDD on an incomplete network. To generate the incomplete network data as the ground truth, we randomly select a certain percentage of nodes as the hidden nodes (Fig. \ref{fig:robust}(a)), and the time series data of these nodes are removed. AUC decreases and MSE increases as the fraction of the number of unobserved nodes increases on both spring and voice dynamics, as shown in Fig. \ref{fig:robust}(c); however, the sensitivity depends on various types of dynamics. It is found that when the proportion of missing nodes reaches 50$\%$, the inference accuracy is still above 95$\%$, which proves that our model can achieve superior results in the absence of normally sufficient amounts of data.

Furthermore, we test the ability of AIDD to reveal unknown network structures of unobservable nodes on CMN and Voter dynamics, with only the number of hidden nodes available. We completed this task by performing the same interaction inference task, setting the states for unknown nodes to random values. Figure \ref{fig:robust}(d) shows the AUCs of the link structures of unknown networks on Voter and CMN dynamics. The results reveal that the network inference accuracy is robust for missing nodes. The algorithm can recover the interactions even for unobservable nodes with over 80\% accuracy. 
See ~\cite{chen2020network} and Supplemental Material at section 5 for more details of the algorithm.

\subsubsection{Reconstruction performance in actual scenarios}
To verify that our algorithm can be applied to actual scenarios and not only on toy models, we attempt to infer the real subnetwork structure from the known transcriptional network of yeast S. cerevisiae according to the time series data of mRNA concentrations generated by GeneNetWeaver (GNW)~\cite{schaffter2011genenetweaver}, a famous simulator for gene dynamics.

The networks used by GNWs are extracted from known biological interaction networks (Escherichia coli, Saccharomyces cerevisiae, etc.). On these networks, GNW uses a set of dynamical equations to simulate the transcription and translation processes, and it has considered many factors close to real situations. Therefore, GNW is a famous platform for benchmarking and performance assessment of network inference methods.

\begin{figure*}[ht!]
\centering
\includegraphics[scale=0.5]{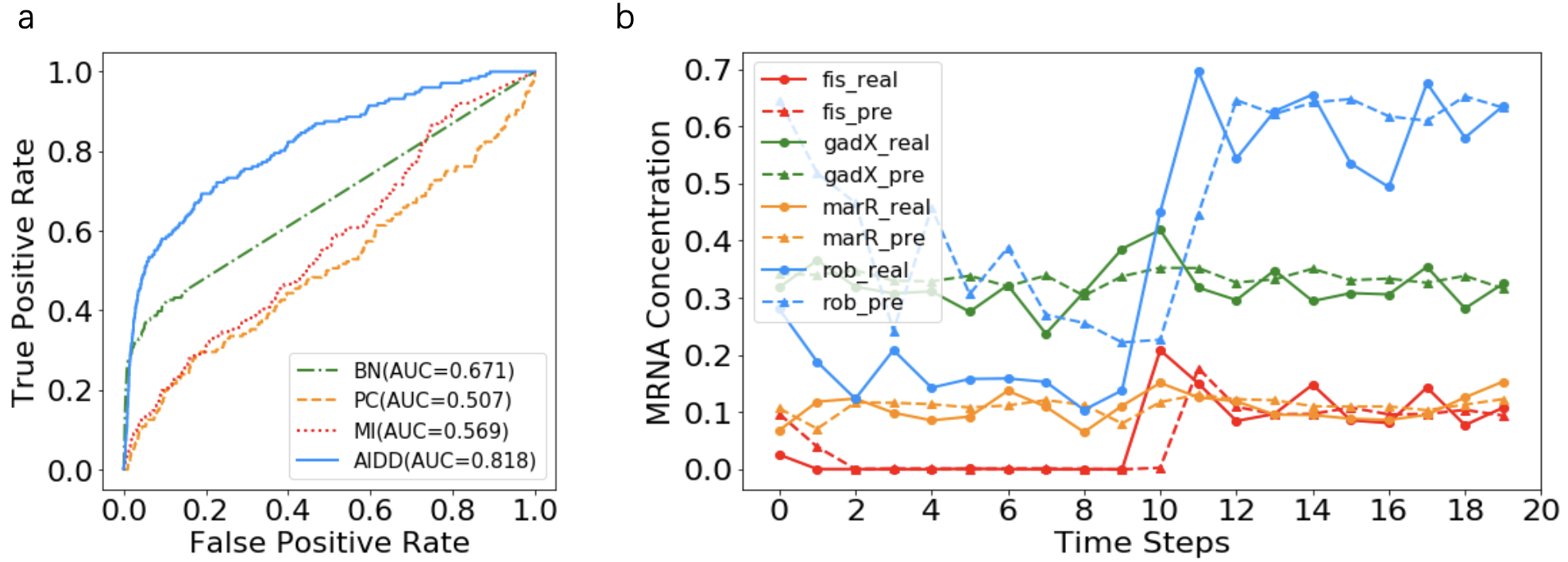}
\caption{Performances of AIDD and other compared methods on network and dynamics reconstruction for the gene regulatory network of yeast S. cerevisiae with 100 nodes. (a) ROC curves of different network inference methods. True positive rate (y-axis) means the fraction of the actual links are classified correctly by our model, and the false positive rate (x-axis) represents the fraction of the node pairs without actual link are classified as links by our model}. The comparison methods include Bayesian network (BN), partial correlation (PC), mutual information (MI), and AIDD. The AUC for different methods is marked in the legend. (b) shows the comparison between the observed time series of the expression data (real) and the predicted data on selected genes. In this plot, the solid lines represent the predictions and the dotted lines represent the observed data. 
\label{fig:gene}
\end{figure*}

In the experiment, we used yeast S. cerevisiae gene network with 100 nodes as the benchmark gene network, and we used the default parameters of DREAM4\_In-Silico in GeneNetWeaver software to generate data. For the dynamics simulator, we use different neural networks for each node because of the heterogeneity of node dynamics and the existence of latent variables, noise, and perturbations~\cite{schaffter2011genenetweaver}. We compare our method with partial correlation, Bayesian network inference, and mutual information algorithms. Our method outperforms others on network inference (Fig. \ref{fig:gene}(a)) on the AUC(0.82). It can also predict the dynamics with a relatively high accuracy (the average absolute error (MAE) is 0.038, see Fig. \ref{fig:gene}. This indicates that our method can perform well realistic gene regulatory dynamics.

\subsubsection{Control Experiments}
To further verify that AIDD has the ability to learn the ground truth dynamics and that the well trained adjacency matrix $\hat{A}$ and node dynamics model $\hat{f_i}(\hat{X}_i^{t+1}|\mathbf{x}^t,\hat{A})$ for all $i$ can replace the original system ($A$ and $f_i$), we design control experiments. The reasons why we choose the control problem as the test bed for our model include (1) the control problem in complex network is very important, and it has relevance to many engineering fields~\cite{liu2016control}; (2) the control problem is more difficult than predictions. This is true because 
to control a system means to intervene in it. As a result, we have stood at least on the second level of the causal hierarchy~\cite{pearl2018book}.

Here, our control problem is to find a control law based on the well trained system (learned network $\hat{A}$ and node dynamics $\hat{f}_i$) such that some control objective can be achieved. And after the optimized control law is found, we can further apply it on the ground truth system (real network structure $A$ and real node dynamics $f_i$). We hope that the same control objective on the ground truth system can also be achieved. It is obvious that only if the learned network $\hat{A}$ and node dynamics $\hat{f_i}$ are very closed to the ground truth system, the same control target can be realized. In this way, we can test if the learned system can reproduce the causal mechanisms of the ground truth system but not merely the pseudo-causal laws based on correlations.

\begin{figure*}[ht!]
\centering
\includegraphics[scale=0.52]{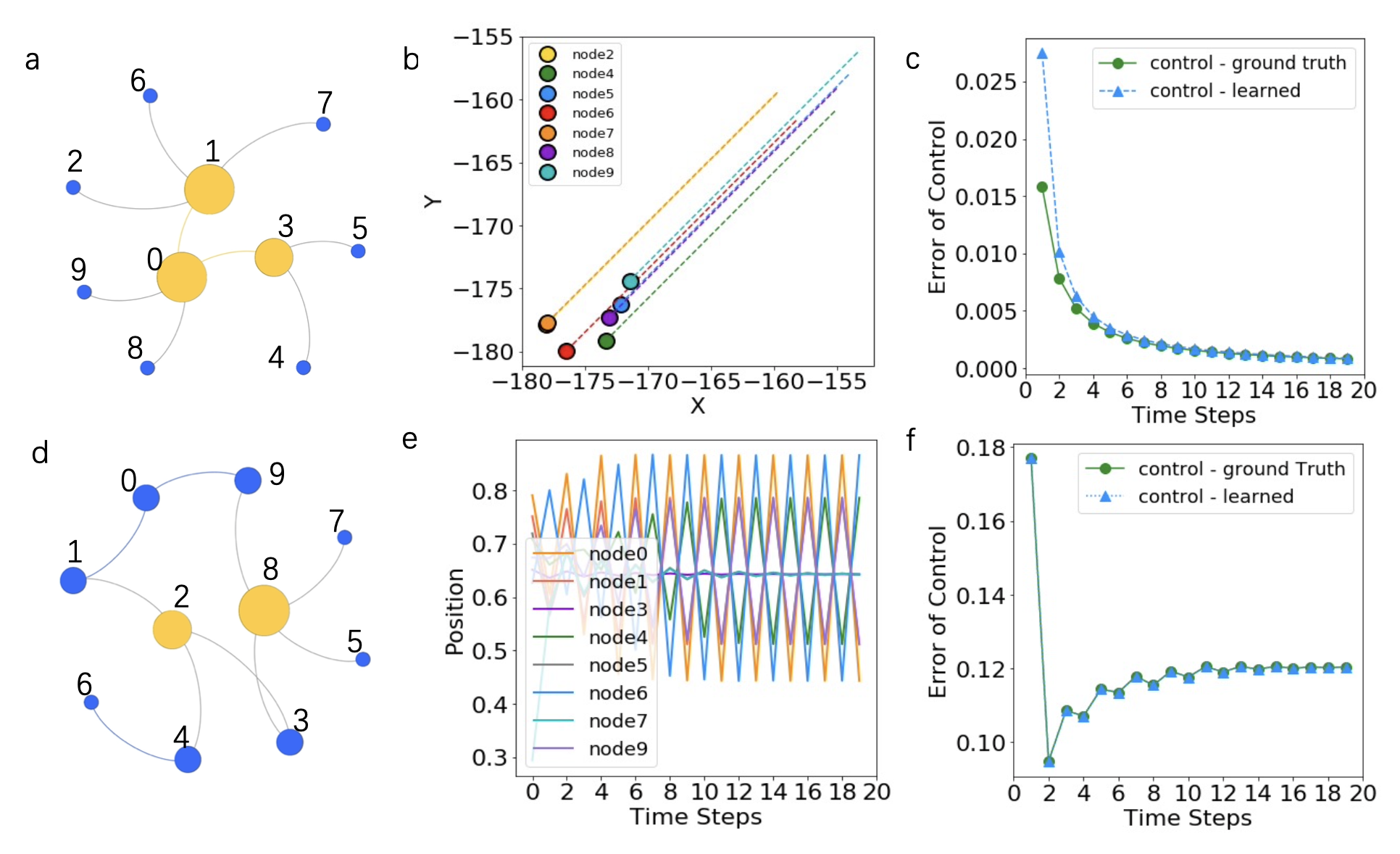}
\caption{The control experiments on learned models. (a) shows the spring network that we studied for the first experiment. Three large nodes are driver nodes, and the others are target nodes. The control objective is to request all masses to have the same movement direction. (b) shows the final movement states of all the target nodes under the controls. $X$ and $Y$ represents the two-dimensional coordinates of positions for all masses.} (c) shows two loss curves for evaluating goal achievement versus time steps of the controls. One represents the results of learned model, and the other is the ground truth. (d) is a coupled mapping network that we studied in the second experiment. Two large nodes were selected as driver nodes. The control objective is to ask all oscillators to have the same value of 0.6, which is the mean of the value range for all nodes. (e) shows the oscillations of all target nodes during control. (f) shows two loss curves for evaluating goal achievement versus time steps of the controls. One is for the trained model, and the other for the ground truth.
\label{fig:control}
\end{figure*}

For example, suppose the ground truth system is a spring-mass dynamical system with 10 nodes (masses) and 18 edges(springs) to form the network $A$ as shown in Figure \ref{fig:control} (a), and the node dynamic $f_i$ is the spring-mass dynamics. We can train an AIDD model ($\hat{A}, \hat{f_i}$) to reconstruct this network and node dynamics. After a large number of steps of training, we can do the control experiment under the trained system ($\hat{A}, \hat{f_i})$. Suppose our control objective is to make all nodes to move in the same direction. However, we can only exert forces on the three driver nodes (0,1,3) as shown in Figure \ref{fig:control} (a). We can train another neural network (the controller) to learn the control law, i.e., how to exert forces on drivers to let all nodes move in the same direction. The inputs to the controller are the states of all nodes on the learned system at each time, the outputs are the forces (including the magnitude and the direction) that will exert on the three drivers at that time, and the training loss function is the error in achieving the control objective. After training enough time steps, when the controller converges, we can transfer the controller optimized on the learned system ($\hat{A},\hat{f_i}$) onto the ground truth system to achieve the same objective. Figure \ref{fig:control} (b) and (c) show the trajectories of nodes and the errors in achieving the objectives on the ground truth system and the learned system at each time step after the controller has exerted forces onto the drivers. The two curves collapse together after about six time steps of controls which means the learned system and the ground truth have similar behaviors under the control law. And the fact that the errors tend to zero means the control objectives have been realized.

The second example is a similar control experiment on CMN dynamics. In which the nodes are oscillators, the links are the connections,  the network is a small world network generated by the WS model, and the node dynamic is the CMN dynamics. The control objective is to synchronize all nodes by exerting forces on the driver nodes (2 and 8). From Fig. \ref{fig:control} (e) and (f), the controls are not well achieved for all nodes because the error curves do not converge to zeros. However, the two error curves overlap very well, indicating that the surrogate behaves identically to the ground truth model. 

These experiments show that the AIDD model can learn the mechanism of causality such that we can apply the control law optimized on the learned system onto the real system.

\section{Conclusion}
In this paper, we propose a unified framework for network structure reconstruction and node dynamics simulation in the same time. We also propose a new standard based on control tasks to evaluate whether the true network dynamics can be learned.

Compared to our previous work in \cite{zhang2019general}, we formulate our problem in a more rigor mathematical form which is based on markov dynamics and localised networked interactions. We also propose a new mathematical framework for our model. A unified objective function based on log-likelihood is provided which is the theoretical foundation of why the model can be applied on different dynamics. We also describe how the optimization objective function can be decomposed into single nodes which is the basis for a new deep learning based node by node network reconstruction method. 
From experimental aspect, the main highlights of the model include scalability, universality, and robustness. The high scalability is reflected by the fact that the model can be applied to large networks with more than thousands of nodes with more than 90\% accuracy because the training procedure can be taken node by node. It is a universal framework because it can be applied to various types of dynamics, including continuous, discrete, and binary. Our model is robust not only on noisy input signals, but also on unobservable nodes. It can recover an entire network even when time series data is missing with more than 90\% accuracy. It was also shown to work well on datasets generated by GeneNetWeaver, which emulates the real environment of gene regulatory network dynamics.

Furthermore, we propose a new method based on controls to test the validity of the learned model. Control experiments on spring and CMN dynamics of small networks have proved that well-trained models can replace the real systems. 

This framework has many potential applications. For example, our method can be used to infer missing links according to the dynamics information. It can also be used in time series forecasting. In contrast to other forecasting models, a clear binary network can be output by the model, which can provide deeper insights into element interactions and potential causal links, increasing the explanability of the model. 

However, some drawbacks are present in the current framework. First, a large amount of training data, especially the time series in diverse initial conditions, is required to obtain a good model. Nevertheless, it is difficult to obtain time series data in many cases. Hence we may develop new models that are suitable for small data. 

Second, all the dynamics considered in this paper are Markovian, but this property is hardly satisfied in real cases. New extensions and experiments on non-Markovian dynamics should be conducted. For example, we can use a recurrent neural network instead of a feed-forward network as the dynamics simulating component.

\begin{acknowledgments}
We acknowledge Prof. Qinghua Chen, Dr. Lifei Wang, and the workshops in Swarma Club for the helpful discussion. We acknowledge the support of the National Natural Science Foundation of China (NSFC) under grant numbers 61673070.
\end{acknowledgments}

\nocite{*}

\bibliography{apssamp}

\begin{thebibliography}{54}%
\makeatletter
\providecommand \@ifxundefined [1]{%
 \@ifx{#1\undefined}
}%
\providecommand \@ifnum [1]{%
 \ifnum #1\expandafter \@firstoftwo
 \else \expandafter \@secondoftwo
 \fi
}%
\providecommand \@ifx [1]{%
 \ifx #1\expandafter \@firstoftwo
 \else \expandafter \@secondoftwo
 \fi
}%
\providecommand \natexlab [1]{#1}%
\providecommand \enquote  [1]{``#1''}%
\providecommand \bibnamefont  [1]{#1}%
\providecommand \bibfnamefont [1]{#1}%
\providecommand \citenamefont [1]{#1}%
\providecommand \href@noop [0]{\@secondoftwo}%
\providecommand \href [0]{\begingroup \@sanitize@url \@href}%
\providecommand \@href[1]{\@@startlink{#1}\@@href}%
\providecommand \@@href[1]{\endgroup#1\@@endlink}%
\providecommand \@sanitize@url [0]{\catcode `\\12\catcode `\$12\catcode
  `\&12\catcode `\#12\catcode `\^12\catcode `\_12\catcode `\%12\relax}%
\providecommand \@@startlink[1]{}%
\providecommand \@@endlink[0]{}%
\providecommand \url  [0]{\begingroup\@sanitize@url \@url }%
\providecommand \@url [1]{\endgroup\@href {#1}{\urlprefix }}%
\providecommand \urlprefix  [0]{URL }%
\providecommand \Eprint [0]{\href }%
\providecommand \doibase [0]{https://doi.org/}%
\providecommand \selectlanguage [0]{\@gobble}%
\providecommand \bibinfo  [0]{\@secondoftwo}%
\providecommand \bibfield  [0]{\@secondoftwo}%
\providecommand \translation [1]{[#1]}%
\providecommand \BibitemOpen [0]{}%
\providecommand \bibitemStop [0]{}%
\providecommand \bibitemNoStop [0]{.\EOS\space}%
\providecommand \EOS [0]{\spacefactor3000\relax}%
\providecommand \BibitemShut  [1]{\csname bibitem#1\endcsname}%
\let\auto@bib@innerbib\@empty
\bibitem [{\citenamefont {Boccaletti}\ \emph {et~al.}(2006)\citenamefont
  {Boccaletti}, \citenamefont {Latora}, \citenamefont {Moreno}, \citenamefont
  {Chavez},\ and\ \citenamefont {Hwang}}]{boccaletti2006complex}%
  \BibitemOpen
  \bibfield  {author} {\bibinfo {author} {\bibfnamefont {S.}~\bibnamefont
  {Boccaletti}}, \bibinfo {author} {\bibfnamefont {V.}~\bibnamefont {Latora}},
  \bibinfo {author} {\bibfnamefont {Y.}~\bibnamefont {Moreno}}, \bibinfo
  {author} {\bibfnamefont {M.}~\bibnamefont {Chavez}},\ and\ \bibinfo {author}
  {\bibfnamefont {D.-U.}\ \bibnamefont {Hwang}},\ }\bibfield  {title} {\bibinfo
  {title} {Complex networks: Structure and dynamics},\ }\href@noop {}
  {\bibfield  {journal} {\bibinfo  {journal} {Physics reports}\ }\textbf
  {\bibinfo {volume} {424}},\ \bibinfo {pages} {175} (\bibinfo {year}
  {2006})}\BibitemShut {NoStop}%
\bibitem [{\citenamefont {Watts}\ and\ \citenamefont
  {Strogatz}(1998)}]{watts1998collective}%
  \BibitemOpen
  \bibfield  {author} {\bibinfo {author} {\bibfnamefont {D.~J.}\ \bibnamefont
  {Watts}}\ and\ \bibinfo {author} {\bibfnamefont {S.~H.}\ \bibnamefont
  {Strogatz}},\ }\bibfield  {title} {\bibinfo {title} {Collective dynamics of
  ‘small-world’networks},\ }\href@noop {} {\bibfield  {journal} {\bibinfo
  {journal} {Nature}\ }\textbf {\bibinfo {volume} {393}},\ \bibinfo {pages}
  {440} (\bibinfo {year} {1998})}\BibitemShut {NoStop}%
\bibitem [{\citenamefont {Runge}\ \emph {et~al.}(2019)\citenamefont {Runge},
  \citenamefont {Nowack}, \citenamefont {Kretschmer}, \citenamefont {Flaxman},\
  and\ \citenamefont {Sejdinovic}}]{runge2019detecting}%
  \BibitemOpen
  \bibfield  {author} {\bibinfo {author} {\bibfnamefont {J.}~\bibnamefont
  {Runge}}, \bibinfo {author} {\bibfnamefont {P.}~\bibnamefont {Nowack}},
  \bibinfo {author} {\bibfnamefont {M.}~\bibnamefont {Kretschmer}}, \bibinfo
  {author} {\bibfnamefont {S.}~\bibnamefont {Flaxman}},\ and\ \bibinfo {author}
  {\bibfnamefont {D.}~\bibnamefont {Sejdinovic}},\ }\bibfield  {title}
  {\bibinfo {title} {Detecting and quantifying causal associations in large
  nonlinear time series datasets},\ }\href@noop {} {\bibfield  {journal}
  {\bibinfo  {journal} {Science Advances}\ }\textbf {\bibinfo {volume} {5}},\
  \bibinfo {pages} {eaau4996} (\bibinfo {year} {2019})}\BibitemShut {NoStop}%
\bibitem [{\citenamefont {Liu}\ and\ \citenamefont
  {Barab{\'a}si}(2016)}]{liu2016control}%
  \BibitemOpen
  \bibfield  {author} {\bibinfo {author} {\bibfnamefont {Y.-Y.}\ \bibnamefont
  {Liu}}\ and\ \bibinfo {author} {\bibfnamefont {A.-L.}\ \bibnamefont
  {Barab{\'a}si}},\ }\bibfield  {title} {\bibinfo {title} {Control principles
  of complex systems},\ }\href@noop {} {\bibfield  {journal} {\bibinfo
  {journal} {Reviews of Modern Physics}\ }\textbf {\bibinfo {volume} {88}},\
  \bibinfo {pages} {035006} (\bibinfo {year} {2016})}\BibitemShut {NoStop}%
\bibitem [{\citenamefont {Wang}\ \emph {et~al.}(2016)\citenamefont {Wang},
  \citenamefont {Lai},\ and\ \citenamefont {Grebogi}}]{wang2016data}%
  \BibitemOpen
  \bibfield  {author} {\bibinfo {author} {\bibfnamefont {W.-X.}\ \bibnamefont
  {Wang}}, \bibinfo {author} {\bibfnamefont {Y.-C.}\ \bibnamefont {Lai}},\ and\
  \bibinfo {author} {\bibfnamefont {C.}~\bibnamefont {Grebogi}},\ }\bibfield
  {title} {\bibinfo {title} {Data based identification and prediction of
  nonlinear and complex dynamical systems},\ }\href@noop {} {\bibfield
  {journal} {\bibinfo  {journal} {Physics Reports}\ }\textbf {\bibinfo {volume}
  {644}},\ \bibinfo {pages} {1} (\bibinfo {year} {2016})}\BibitemShut {NoStop}%
\bibitem [{\citenamefont {Ha}\ and\ \citenamefont {Jeong}(2020)}]{ha2020deep}%
  \BibitemOpen
  \bibfield  {author} {\bibinfo {author} {\bibfnamefont {S.}~\bibnamefont
  {Ha}}\ and\ \bibinfo {author} {\bibfnamefont {H.}~\bibnamefont {Jeong}},\
  }\bibfield  {title} {\bibinfo {title} {Deep learning reveals hidden
  interactions in complex systems},\ }\href@noop {} {\bibfield  {journal}
  {\bibinfo  {journal} {arXiv preprint arXiv:2001.02539}\ } (\bibinfo {year}
  {2020})}\BibitemShut {NoStop}%
\bibitem [{\citenamefont {Pearl}(2009)}]{pearl2009causality}%
  \BibitemOpen
  \bibfield  {author} {\bibinfo {author} {\bibfnamefont {J.}~\bibnamefont
  {Pearl}},\ }\href@noop {} {\emph {\bibinfo {title} {Causality}}}\ (\bibinfo
  {publisher} {Cambridge university press},\ \bibinfo {year}
  {2009})\BibitemShut {NoStop}%
\bibitem [{\citenamefont {Tank}\ \emph {et~al.}(2018)\citenamefont {Tank},
  \citenamefont {Covert}, \citenamefont {Foti}, \citenamefont {Shojaie},\ and\
  \citenamefont {Fox}}]{tank2018neural}%
  \BibitemOpen
  \bibfield  {author} {\bibinfo {author} {\bibfnamefont {A.}~\bibnamefont
  {Tank}}, \bibinfo {author} {\bibfnamefont {I.}~\bibnamefont {Covert}},
  \bibinfo {author} {\bibfnamefont {N.}~\bibnamefont {Foti}}, \bibinfo {author}
  {\bibfnamefont {A.}~\bibnamefont {Shojaie}},\ and\ \bibinfo {author}
  {\bibfnamefont {E.}~\bibnamefont {Fox}},\ }\bibfield  {title} {\bibinfo
  {title} {Neural granger causality for nonlinear time series},\ }\href@noop {}
  {\bibfield  {journal} {\bibinfo  {journal} {arXiv preprint arXiv:1802.05842}\
  } (\bibinfo {year} {2018})}\BibitemShut {NoStop}%
\bibitem [{\citenamefont {L{\"o}we}\ \emph {et~al.}(2020)\citenamefont
  {L{\"o}we}, \citenamefont {Madras}, \citenamefont {Zemel},\ and\
  \citenamefont {Welling}}]{lowe2020amortized}%
  \BibitemOpen
  \bibfield  {author} {\bibinfo {author} {\bibfnamefont {S.}~\bibnamefont
  {L{\"o}we}}, \bibinfo {author} {\bibfnamefont {D.}~\bibnamefont {Madras}},
  \bibinfo {author} {\bibfnamefont {R.}~\bibnamefont {Zemel}},\ and\ \bibinfo
  {author} {\bibfnamefont {M.}~\bibnamefont {Welling}},\ }\bibfield  {title}
  {\bibinfo {title} {Amortized causal discovery: Learning to infer causal
  graphs from time-series data},\ }\href@noop {} {\bibfield  {journal}
  {\bibinfo  {journal} {arXiv preprint arXiv:2006.10833}\ } (\bibinfo {year}
  {2020})}\BibitemShut {NoStop}%
\bibitem [{\citenamefont {Glymour}\ \emph {et~al.}(2019)\citenamefont
  {Glymour}, \citenamefont {Zhang},\ and\ \citenamefont
  {Spirtes}}]{glymour2019review}%
  \BibitemOpen
  \bibfield  {author} {\bibinfo {author} {\bibfnamefont {C.}~\bibnamefont
  {Glymour}}, \bibinfo {author} {\bibfnamefont {K.}~\bibnamefont {Zhang}},\
  and\ \bibinfo {author} {\bibfnamefont {P.}~\bibnamefont {Spirtes}},\
  }\bibfield  {title} {\bibinfo {title} {Review of causal discovery methods
  based on graphical models},\ }\href@noop {} {\bibfield  {journal} {\bibinfo
  {journal} {Frontiers in genetics}\ }\textbf {\bibinfo {volume} {10}},\
  \bibinfo {pages} {524} (\bibinfo {year} {2019})}\BibitemShut {NoStop}%
\bibitem [{\citenamefont {Peng}\ \emph {et~al.}(2009)\citenamefont {Peng},
  \citenamefont {Wang}, \citenamefont {Zhou},\ and\ \citenamefont
  {Zhu}}]{peng2009partial}%
  \BibitemOpen
  \bibfield  {author} {\bibinfo {author} {\bibfnamefont {J.}~\bibnamefont
  {Peng}}, \bibinfo {author} {\bibfnamefont {P.}~\bibnamefont {Wang}}, \bibinfo
  {author} {\bibfnamefont {N.}~\bibnamefont {Zhou}},\ and\ \bibinfo {author}
  {\bibfnamefont {J.}~\bibnamefont {Zhu}},\ }\bibfield  {title} {\bibinfo
  {title} {Partial correlation estimation by joint sparse regression models},\
  }\href@noop {} {\bibfield  {journal} {\bibinfo  {journal} {Journal of the
  American Statistical Association}\ }\textbf {\bibinfo {volume} {104}},\
  \bibinfo {pages} {735} (\bibinfo {year} {2009})}\BibitemShut {NoStop}%
\bibitem [{\citenamefont {Nitzan}\ \emph {et~al.}(2017)\citenamefont {Nitzan},
  \citenamefont {Casadiego},\ and\ \citenamefont
  {Timme}}]{nitzan2017revealing}%
  \BibitemOpen
  \bibfield  {author} {\bibinfo {author} {\bibfnamefont {M.}~\bibnamefont
  {Nitzan}}, \bibinfo {author} {\bibfnamefont {J.}~\bibnamefont {Casadiego}},\
  and\ \bibinfo {author} {\bibfnamefont {M.}~\bibnamefont {Timme}},\ }\bibfield
   {title} {\bibinfo {title} {Revealing physical interaction networks from
  statistics of collective dynamics},\ }\href@noop {} {\bibfield  {journal}
  {\bibinfo  {journal} {Science advances}\ }\textbf {\bibinfo {volume} {3}},\
  \bibinfo {pages} {e1600396} (\bibinfo {year} {2017})}\BibitemShut {NoStop}%
\bibitem [{\citenamefont {Liu}\ \emph {et~al.}(2018)\citenamefont {Liu},
  \citenamefont {Kim},\ and\ \citenamefont {Shlizerman}}]{liu2018functional}%
  \BibitemOpen
  \bibfield  {author} {\bibinfo {author} {\bibfnamefont {H.}~\bibnamefont
  {Liu}}, \bibinfo {author} {\bibfnamefont {J.}~\bibnamefont {Kim}},\ and\
  \bibinfo {author} {\bibfnamefont {E.}~\bibnamefont {Shlizerman}},\ }\bibfield
   {title} {\bibinfo {title} {Functional connectomics from neural dynamics:
  probabilistic graphical models for neuronal network of caenorhabditis
  elegans},\ }\href@noop {} {\bibfield  {journal} {\bibinfo  {journal}
  {Philosophical Transactions of the Royal Society B: Biological Sciences}\
  }\textbf {\bibinfo {volume} {373}},\ \bibinfo {pages} {20170377} (\bibinfo
  {year} {2018})}\BibitemShut {NoStop}%
\bibitem [{\citenamefont {Runge}(2018)}]{runge2018causal}%
  \BibitemOpen
  \bibfield  {author} {\bibinfo {author} {\bibfnamefont {J.}~\bibnamefont
  {Runge}},\ }\bibfield  {title} {\bibinfo {title} {Causal network
  reconstruction from time series: From theoretical assumptions to practical
  estimation},\ }\href@noop {} {\bibfield  {journal} {\bibinfo  {journal}
  {Chaos: An Interdisciplinary Journal of Nonlinear Science}\ }\textbf
  {\bibinfo {volume} {28}},\ \bibinfo {pages} {075310} (\bibinfo {year}
  {2018})}\BibitemShut {NoStop}%
\bibitem [{\citenamefont {Casadiego}\ \emph {et~al.}(2017)\citenamefont
  {Casadiego}, \citenamefont {Nitzan}, \citenamefont {Hallerberg},\ and\
  \citenamefont {Timme}}]{casadiego2017model}%
  \BibitemOpen
  \bibfield  {author} {\bibinfo {author} {\bibfnamefont {J.}~\bibnamefont
  {Casadiego}}, \bibinfo {author} {\bibfnamefont {M.}~\bibnamefont {Nitzan}},
  \bibinfo {author} {\bibfnamefont {S.}~\bibnamefont {Hallerberg}},\ and\
  \bibinfo {author} {\bibfnamefont {M.}~\bibnamefont {Timme}},\ }\bibfield
  {title} {\bibinfo {title} {Model-free inference of direct network
  interactions from nonlinear collective dynamics},\ }\href@noop {} {\bibfield
  {journal} {\bibinfo  {journal} {Nature communications}\ }\textbf {\bibinfo
  {volume} {8}},\ \bibinfo {pages} {1} (\bibinfo {year} {2017})}\BibitemShut
  {NoStop}%
\bibitem [{\citenamefont {Li}\ \emph {et~al.}(2017)\citenamefont {Li},
  \citenamefont {Xu}, \citenamefont {Peng}, \citenamefont {Kurths},\ and\
  \citenamefont {Yang}}]{li2017reconstruction}%
  \BibitemOpen
  \bibfield  {author} {\bibinfo {author} {\bibfnamefont {L.}~\bibnamefont
  {Li}}, \bibinfo {author} {\bibfnamefont {D.}~\bibnamefont {Xu}}, \bibinfo
  {author} {\bibfnamefont {H.}~\bibnamefont {Peng}}, \bibinfo {author}
  {\bibfnamefont {J.}~\bibnamefont {Kurths}},\ and\ \bibinfo {author}
  {\bibfnamefont {Y.}~\bibnamefont {Yang}},\ }\bibfield  {title} {\bibinfo
  {title} {Reconstruction of complex network based on the noise via {QR}
  decomposition and compressed sensing},\ }\href@noop {} {\bibfield  {journal}
  {\bibinfo  {journal} {Scientific reports}\ }\textbf {\bibinfo {volume} {7}},\
  \bibinfo {pages} {1} (\bibinfo {year} {2017})}\BibitemShut {NoStop}%
\bibitem [{\citenamefont {Granger}(1969)}]{granger1969investigating}%
  \BibitemOpen
  \bibfield  {author} {\bibinfo {author} {\bibfnamefont {C.~W.}\ \bibnamefont
  {Granger}},\ }\bibfield  {title} {\bibinfo {title} {Investigating causal
  relations by econometric models and cross-spectral methods},\ }\href@noop {}
  {\bibfield  {journal} {\bibinfo  {journal} {Econometrica: journal of the
  Econometric Society}\ ,\ \bibinfo {pages} {424}} (\bibinfo {year}
  {1969})}\BibitemShut {NoStop}%
\bibitem [{\citenamefont {Takens}(1981)}]{takens1981detecting}%
  \BibitemOpen
  \bibfield  {author} {\bibinfo {author} {\bibfnamefont {F.}~\bibnamefont
  {Takens}},\ }\bibfield  {title} {\bibinfo {title} {Detecting strange
  attractors in turbulence},\ }in\ \href@noop {} {\emph {\bibinfo {booktitle}
  {Dynamical systems and turbulence, Warwick 1980}}}\ (\bibinfo  {publisher}
  {Springer},\ \bibinfo {year} {1981})\ pp.\ \bibinfo {pages}
  {366--381}\BibitemShut {NoStop}%
\bibitem [{\citenamefont {Brockwell}\ \emph {et~al.}(1991)\citenamefont
  {Brockwell}, \citenamefont {Davis},\ and\ \citenamefont
  {Fienberg}}]{brockwell1991time}%
  \BibitemOpen
  \bibfield  {author} {\bibinfo {author} {\bibfnamefont {P.~J.}\ \bibnamefont
  {Brockwell}}, \bibinfo {author} {\bibfnamefont {R.~A.}\ \bibnamefont
  {Davis}},\ and\ \bibinfo {author} {\bibfnamefont {S.~E.}\ \bibnamefont
  {Fienberg}},\ }\href@noop {} {\emph {\bibinfo {title} {Time series: theory
  and methods}}}\ (\bibinfo  {publisher} {Springer Science \& Business Media},\
  \bibinfo {year} {1991})\BibitemShut {NoStop}%
\bibitem [{\citenamefont {Jaeger}(2001)}]{jaeger2001echo}%
  \BibitemOpen
  \bibfield  {author} {\bibinfo {author} {\bibfnamefont {H.}~\bibnamefont
  {Jaeger}},\ }\bibfield  {title} {\bibinfo {title} {The “echo state”
  approach to analysing and training recurrent neural networks-with an erratum
  note},\ }\href@noop {} {\bibfield  {journal} {\bibinfo  {journal} {Bonn,
  Germany: German National Research Center for Information Technology GMD
  Technical Report}\ }\textbf {\bibinfo {volume} {148}},\ \bibinfo {pages} {13}
  (\bibinfo {year} {2001})}\BibitemShut {NoStop}%
\bibitem [{\citenamefont {Schrauwen}\ \emph {et~al.}(2007)\citenamefont
  {Schrauwen}, \citenamefont {Verstraeten},\ and\ \citenamefont
  {Van~Campenhout}}]{schrauwen2007overview}%
  \BibitemOpen
  \bibfield  {author} {\bibinfo {author} {\bibfnamefont {B.}~\bibnamefont
  {Schrauwen}}, \bibinfo {author} {\bibfnamefont {D.}~\bibnamefont
  {Verstraeten}},\ and\ \bibinfo {author} {\bibfnamefont {J.}~\bibnamefont
  {Van~Campenhout}},\ }\bibfield  {title} {\bibinfo {title} {An overview of
  reservoir computing: theory, applications and implementations},\ }in\
  \href@noop {} {\emph {\bibinfo {booktitle} {Proceedings of the 15th european
  symposium on artificial neural networks. p. 471-482 2007}}}\ (\bibinfo {year}
  {2007})\ pp.\ \bibinfo {pages} {471--482}\BibitemShut {NoStop}%
\bibitem [{\citenamefont {Scarselli}\ \emph {et~al.}(2008)\citenamefont
  {Scarselli}, \citenamefont {Gori}, \citenamefont {Tsoi}, \citenamefont
  {Hagenbuchner},\ and\ \citenamefont {Monfardini}}]{scarselli2008graph}%
  \BibitemOpen
  \bibfield  {author} {\bibinfo {author} {\bibfnamefont {F.}~\bibnamefont
  {Scarselli}}, \bibinfo {author} {\bibfnamefont {M.}~\bibnamefont {Gori}},
  \bibinfo {author} {\bibfnamefont {A.~C.}\ \bibnamefont {Tsoi}}, \bibinfo
  {author} {\bibfnamefont {M.}~\bibnamefont {Hagenbuchner}},\ and\ \bibinfo
  {author} {\bibfnamefont {G.}~\bibnamefont {Monfardini}},\ }\bibfield  {title}
  {\bibinfo {title} {The graph neural network model},\ }\href@noop {}
  {\bibfield  {journal} {\bibinfo  {journal} {IEEE transactions on neural
  networks}\ }\textbf {\bibinfo {volume} {20}},\ \bibinfo {pages} {61}
  (\bibinfo {year} {2008})}\BibitemShut {NoStop}%
\bibitem [{\citenamefont {Battaglia}\ \emph {et~al.}(2018)\citenamefont
  {Battaglia}, \citenamefont {Hamrick}, \citenamefont {Bapst}, \citenamefont
  {Sanchez-Gonzalez}, \citenamefont {Zambaldi}, \citenamefont {Malinowski},
  \citenamefont {Tacchetti}, \citenamefont {Raposo}, \citenamefont {Santoro},
  \citenamefont {Faulkner} \emph {et~al.}}]{battaglia2018relational}%
  \BibitemOpen
  \bibfield  {author} {\bibinfo {author} {\bibfnamefont {P.~W.}\ \bibnamefont
  {Battaglia}}, \bibinfo {author} {\bibfnamefont {J.~B.}\ \bibnamefont
  {Hamrick}}, \bibinfo {author} {\bibfnamefont {V.}~\bibnamefont {Bapst}},
  \bibinfo {author} {\bibfnamefont {A.}~\bibnamefont {Sanchez-Gonzalez}},
  \bibinfo {author} {\bibfnamefont {V.}~\bibnamefont {Zambaldi}}, \bibinfo
  {author} {\bibfnamefont {M.}~\bibnamefont {Malinowski}}, \bibinfo {author}
  {\bibfnamefont {A.}~\bibnamefont {Tacchetti}}, \bibinfo {author}
  {\bibfnamefont {D.}~\bibnamefont {Raposo}}, \bibinfo {author} {\bibfnamefont
  {A.}~\bibnamefont {Santoro}}, \bibinfo {author} {\bibfnamefont
  {R.}~\bibnamefont {Faulkner}}, \emph {et~al.},\ }\bibfield  {title} {\bibinfo
  {title} {Relational inductive biases, deep learning, and graph networks},\
  }\href@noop {} {\bibfield  {journal} {\bibinfo  {journal} {arXiv preprint
  arXiv:1806.01261}\ } (\bibinfo {year} {2018})}\BibitemShut {NoStop}%
\bibitem [{\citenamefont {{Wu}}\ \emph {et~al.}(2021)\citenamefont {{Wu}},
  \citenamefont {{Pan}}, \citenamefont {{Chen}}, \citenamefont {{Long}},
  \citenamefont {{Zhang}},\ and\ \citenamefont {{Yu}}}]{wu2020comprehensive}%
  \BibitemOpen
  \bibfield  {author} {\bibinfo {author} {\bibfnamefont {Z.}~\bibnamefont
  {{Wu}}}, \bibinfo {author} {\bibfnamefont {S.}~\bibnamefont {{Pan}}},
  \bibinfo {author} {\bibfnamefont {F.}~\bibnamefont {{Chen}}}, \bibinfo
  {author} {\bibfnamefont {G.}~\bibnamefont {{Long}}}, \bibinfo {author}
  {\bibfnamefont {C.}~\bibnamefont {{Zhang}}},\ and\ \bibinfo {author}
  {\bibfnamefont {P.~S.}\ \bibnamefont {{Yu}}},\ }\bibfield  {title} {\bibinfo
  {title} {A comprehensive survey on graph neural networks},\ }\href@noop {}
  {\bibfield  {journal} {\bibinfo  {journal} {IEEE Transactions on Neural
  Networks and Learning Systems}\ }\textbf {\bibinfo {volume} {32}},\ \bibinfo
  {pages} {4} (\bibinfo {year} {2021})}\BibitemShut {NoStop}%
\bibitem [{\citenamefont {Sanchez-Gonzalez}\ \emph {et~al.}(2018)\citenamefont
  {Sanchez-Gonzalez}, \citenamefont {Heess}, \citenamefont {Springenberg},
  \citenamefont {Merel}, \citenamefont {Riedmiller}, \citenamefont {Hadsell},\
  and\ \citenamefont {Battaglia}}]{sanchez2018graph}%
  \BibitemOpen
  \bibfield  {author} {\bibinfo {author} {\bibfnamefont {A.}~\bibnamefont
  {Sanchez-Gonzalez}}, \bibinfo {author} {\bibfnamefont {N.}~\bibnamefont
  {Heess}}, \bibinfo {author} {\bibfnamefont {J.~T.}\ \bibnamefont
  {Springenberg}}, \bibinfo {author} {\bibfnamefont {J.}~\bibnamefont {Merel}},
  \bibinfo {author} {\bibfnamefont {M.}~\bibnamefont {Riedmiller}}, \bibinfo
  {author} {\bibfnamefont {R.}~\bibnamefont {Hadsell}},\ and\ \bibinfo {author}
  {\bibfnamefont {P.}~\bibnamefont {Battaglia}},\ }\bibfield  {title} {\bibinfo
  {title} {Graph networks as learnable physics engines for inference and
  control},\ }in\ \href@noop {} {\emph {\bibinfo {booktitle} {International
  Conference on Machine Learning}}}\ (\bibinfo {organization} {PMLR},\ \bibinfo
  {year} {2018})\ pp.\ \bibinfo {pages} {4470--4479}\BibitemShut {NoStop}%
\bibitem [{\citenamefont {Kipf}\ \emph {et~al.}(2018)\citenamefont {Kipf},
  \citenamefont {Fetaya}, \citenamefont {Wang}, \citenamefont {Welling},\ and\
  \citenamefont {Zemel}}]{kipf2018neural}%
  \BibitemOpen
  \bibfield  {author} {\bibinfo {author} {\bibfnamefont {T.}~\bibnamefont
  {Kipf}}, \bibinfo {author} {\bibfnamefont {E.}~\bibnamefont {Fetaya}},
  \bibinfo {author} {\bibfnamefont {K.-C.}\ \bibnamefont {Wang}}, \bibinfo
  {author} {\bibfnamefont {M.}~\bibnamefont {Welling}},\ and\ \bibinfo {author}
  {\bibfnamefont {R.}~\bibnamefont {Zemel}},\ }\bibfield  {title} {\bibinfo
  {title} {Neural relational inference for interacting systems},\ }in\
  \href@noop {} {\emph {\bibinfo {booktitle} {International Conference on
  Machine Learning}}}\ (\bibinfo {organization} {PMLR},\ \bibinfo {year}
  {2018})\ pp.\ \bibinfo {pages} {2688--2697}\BibitemShut {NoStop}%
\bibitem [{\citenamefont {Zhang}\ \emph
  {et~al.}(2019{\natexlab{a}})\citenamefont {Zhang}, \citenamefont {Wang},
  \citenamefont {Wang}, \citenamefont {Tao}, \citenamefont {Xiao},
  \citenamefont {Mou}, \citenamefont {Cai},\ and\ \citenamefont
  {Zhang}}]{zhang2019neural}%
  \BibitemOpen
  \bibfield  {author} {\bibinfo {author} {\bibfnamefont {Z.}~\bibnamefont
  {Zhang}}, \bibinfo {author} {\bibfnamefont {L.}~\bibnamefont {Wang}},
  \bibinfo {author} {\bibfnamefont {S.}~\bibnamefont {Wang}}, \bibinfo {author}
  {\bibfnamefont {R.}~\bibnamefont {Tao}}, \bibinfo {author} {\bibfnamefont
  {J.}~\bibnamefont {Xiao}}, \bibinfo {author} {\bibfnamefont {M.}~\bibnamefont
  {Mou}}, \bibinfo {author} {\bibfnamefont {J.}~\bibnamefont {Cai}},\ and\
  \bibinfo {author} {\bibfnamefont {J.}~\bibnamefont {Zhang}},\ }\bibfield
  {title} {\bibinfo {title} {Neural gene network constructor: A neural based
  model for reconstructing gene regulatory network},\ }\href@noop {} {\bibfield
   {journal} {\bibinfo  {journal} {bioRxiv}\ ,\ \bibinfo {pages} {842369}}
  (\bibinfo {year} {2019}{\natexlab{a}})}\BibitemShut {NoStop}%
\bibitem [{\citenamefont {Zheng}\ \emph {et~al.}(2020)\citenamefont {Zheng},
  \citenamefont {Fan}, \citenamefont {Wang},\ and\ \citenamefont
  {Qi}}]{zheng2020gman}%
  \BibitemOpen
  \bibfield  {author} {\bibinfo {author} {\bibfnamefont {C.}~\bibnamefont
  {Zheng}}, \bibinfo {author} {\bibfnamefont {X.}~\bibnamefont {Fan}}, \bibinfo
  {author} {\bibfnamefont {C.}~\bibnamefont {Wang}},\ and\ \bibinfo {author}
  {\bibfnamefont {J.}~\bibnamefont {Qi}},\ }\bibfield  {title} {\bibinfo
  {title} {Gman: A graph multi-attention network for traffic prediction},\ }in\
  \href@noop {} {\emph {\bibinfo {booktitle} {Proceedings of the AAAI
  Conference on Artificial Intelligence}}},\ Vol.~\bibinfo {volume} {34}\
  (\bibinfo {year} {2020})\ pp.\ \bibinfo {pages} {1234--1241}\BibitemShut
  {NoStop}%
\bibitem [{\citenamefont {Alet}\ \emph {et~al.}(2019)\citenamefont {Alet},
  \citenamefont {Weng}, \citenamefont {Lozano-P{\'e}rez},\ and\ \citenamefont
  {Kaelbling}}]{alet2019neural}%
  \BibitemOpen
  \bibfield  {author} {\bibinfo {author} {\bibfnamefont {F.}~\bibnamefont
  {Alet}}, \bibinfo {author} {\bibfnamefont {E.}~\bibnamefont {Weng}}, \bibinfo
  {author} {\bibfnamefont {T.}~\bibnamefont {Lozano-P{\'e}rez}},\ and\ \bibinfo
  {author} {\bibfnamefont {L.~P.}\ \bibnamefont {Kaelbling}},\ }\bibfield
  {title} {\bibinfo {title} {Neural relational inference with fast modular
  meta-learning},\ }\href@noop {} {\bibfield  {journal} {\bibinfo  {journal}
  {Advances in Neural Information Processing Systems}\ }\textbf {\bibinfo
  {volume} {32}},\ \bibinfo {pages} {11827} (\bibinfo {year}
  {2019})}\BibitemShut {NoStop}%
\bibitem [{\citenamefont {Zhang}\ \emph
  {et~al.}(2019{\natexlab{b}})\citenamefont {Zhang}, \citenamefont {Zhao},
  \citenamefont {Liu}, \citenamefont {Wang}, \citenamefont {Tao}, \citenamefont
  {Xin},\ and\ \citenamefont {Zhang}}]{zhang2019general}%
  \BibitemOpen
  \bibfield  {author} {\bibinfo {author} {\bibfnamefont {Z.}~\bibnamefont
  {Zhang}}, \bibinfo {author} {\bibfnamefont {Y.}~\bibnamefont {Zhao}},
  \bibinfo {author} {\bibfnamefont {J.}~\bibnamefont {Liu}}, \bibinfo {author}
  {\bibfnamefont {S.}~\bibnamefont {Wang}}, \bibinfo {author} {\bibfnamefont
  {R.}~\bibnamefont {Tao}}, \bibinfo {author} {\bibfnamefont {R.}~\bibnamefont
  {Xin}},\ and\ \bibinfo {author} {\bibfnamefont {J.}~\bibnamefont {Zhang}},\
  }\bibfield  {title} {\bibinfo {title} {A general deep learning framework for
  network reconstruction and dynamics learning},\ }\href@noop {} {\bibfield
  {journal} {\bibinfo  {journal} {Applied Network Science}\ }\textbf {\bibinfo
  {volume} {4}},\ \bibinfo {pages} {1} (\bibinfo {year}
  {2019}{\natexlab{b}})}\BibitemShut {NoStop}%
\bibitem [{\citenamefont {Pareja}\ \emph {et~al.}(2020)\citenamefont {Pareja},
  \citenamefont {Domeniconi}, \citenamefont {Chen}, \citenamefont {Ma},
  \citenamefont {Suzumura}, \citenamefont {Kanezashi}, \citenamefont {Kaler},
  \citenamefont {Schardl},\ and\ \citenamefont
  {Leiserson}}]{pareja2020evolvegcn}%
  \BibitemOpen
  \bibfield  {author} {\bibinfo {author} {\bibfnamefont {A.}~\bibnamefont
  {Pareja}}, \bibinfo {author} {\bibfnamefont {G.}~\bibnamefont {Domeniconi}},
  \bibinfo {author} {\bibfnamefont {J.}~\bibnamefont {Chen}}, \bibinfo {author}
  {\bibfnamefont {T.}~\bibnamefont {Ma}}, \bibinfo {author} {\bibfnamefont
  {T.}~\bibnamefont {Suzumura}}, \bibinfo {author} {\bibfnamefont
  {H.}~\bibnamefont {Kanezashi}}, \bibinfo {author} {\bibfnamefont
  {T.}~\bibnamefont {Kaler}}, \bibinfo {author} {\bibfnamefont
  {T.}~\bibnamefont {Schardl}},\ and\ \bibinfo {author} {\bibfnamefont
  {C.}~\bibnamefont {Leiserson}},\ }\bibfield  {title} {\bibinfo {title}
  {Evolvegcn: Evolving graph convolutional networks for dynamic graphs},\ }in\
  \href@noop {} {\emph {\bibinfo {booktitle} {Proceedings of the AAAI
  Conference on Artificial Intelligence}}},\ Vol.~\bibinfo {volume} {34}\
  (\bibinfo {year} {2020})\ pp.\ \bibinfo {pages} {5363--5370}\BibitemShut
  {NoStop}%
\bibitem [{\citenamefont {Veličković}\ \emph {et~al.}(2018)\citenamefont
  {Veličković}, \citenamefont {Cucurull}, \citenamefont {Casanova},
  \citenamefont {Romero}, \citenamefont {Liò},\ and\ \citenamefont
  {Bengio}}]{velivckovic2017graph}%
  \BibitemOpen
  \bibfield  {author} {\bibinfo {author} {\bibfnamefont {P.}~\bibnamefont
  {Veličković}}, \bibinfo {author} {\bibfnamefont {G.}~\bibnamefont
  {Cucurull}}, \bibinfo {author} {\bibfnamefont {A.}~\bibnamefont {Casanova}},
  \bibinfo {author} {\bibfnamefont {A.}~\bibnamefont {Romero}}, \bibinfo
  {author} {\bibfnamefont {P.}~\bibnamefont {Liò}},\ and\ \bibinfo {author}
  {\bibfnamefont {Y.}~\bibnamefont {Bengio}},\ }\bibfield  {title} {\bibinfo
  {title} {Graph attention networks},\ }in\ \href@noop {} {\emph {\bibinfo
  {booktitle} {International Conference on Learning Representations}}}\
  (\bibinfo  {publisher} {OpenReview.net},\ \bibinfo {year} {2018})\BibitemShut
  {NoStop}%
\bibitem [{\citenamefont {Wang}\ \emph {et~al.}(2020)\citenamefont {Wang},
  \citenamefont {Zhang}, \citenamefont {Ma}, \citenamefont {Zhao},
  \citenamefont {Jiang}, \citenamefont {Chawla},\ and\ \citenamefont
  {Jiang}}]{wang2020learning}%
  \BibitemOpen
  \bibfield  {author} {\bibinfo {author} {\bibfnamefont {D.}~\bibnamefont
  {Wang}}, \bibinfo {author} {\bibfnamefont {Z.}~\bibnamefont {Zhang}},
  \bibinfo {author} {\bibfnamefont {Y.}~\bibnamefont {Ma}}, \bibinfo {author}
  {\bibfnamefont {T.}~\bibnamefont {Zhao}}, \bibinfo {author} {\bibfnamefont
  {T.}~\bibnamefont {Jiang}}, \bibinfo {author} {\bibfnamefont {N.~V.}\
  \bibnamefont {Chawla}},\ and\ \bibinfo {author} {\bibfnamefont
  {M.}~\bibnamefont {Jiang}},\ }\bibfield  {title} {\bibinfo {title} {Learning
  attribute-structure co-evolutions in dynamic graphs},\ }\href@noop {}
  {\bibfield  {journal} {\bibinfo  {journal} {arXiv preprint arXiv:2007.13004}\
  } (\bibinfo {year} {2020})}\BibitemShut {NoStop}%
\bibitem [{\citenamefont {Wu}\ \emph {et~al.}(2019)\citenamefont {Wu},
  \citenamefont {Pan}, \citenamefont {Long}, \citenamefont {Jiang},\ and\
  \citenamefont {Zhang}}]{ijcai2019-264}%
  \BibitemOpen
  \bibfield  {author} {\bibinfo {author} {\bibfnamefont {Z.}~\bibnamefont
  {Wu}}, \bibinfo {author} {\bibfnamefont {S.}~\bibnamefont {Pan}}, \bibinfo
  {author} {\bibfnamefont {G.}~\bibnamefont {Long}}, \bibinfo {author}
  {\bibfnamefont {J.}~\bibnamefont {Jiang}},\ and\ \bibinfo {author}
  {\bibfnamefont {C.}~\bibnamefont {Zhang}},\ }\bibfield  {title} {\bibinfo
  {title} {Graph wavenet for deep spatial-temporal graph modeling},\ }in\
  \href@noop {} {\emph {\bibinfo {booktitle} {Proceedings of the Twenty-Eighth
  International Joint Conference on Artificial Intelligence, {IJCAI-19}}}}\
  (\bibinfo  {publisher} {International Joint Conferences on Artificial
  Intelligence Organization},\ \bibinfo {year} {2019})\ pp.\ \bibinfo {pages}
  {1907--1913}\BibitemShut {NoStop}%
\bibitem [{\citenamefont {Ayed}\ \emph {et~al.}(2019)\citenamefont {Ayed},
  \citenamefont {de~B{\'e}zenac}, \citenamefont {Pajot}, \citenamefont
  {Brajard},\ and\ \citenamefont {Gallinari}}]{ayed2019learning}%
  \BibitemOpen
  \bibfield  {author} {\bibinfo {author} {\bibfnamefont {I.}~\bibnamefont
  {Ayed}}, \bibinfo {author} {\bibfnamefont {E.}~\bibnamefont
  {de~B{\'e}zenac}}, \bibinfo {author} {\bibfnamefont {A.}~\bibnamefont
  {Pajot}}, \bibinfo {author} {\bibfnamefont {J.}~\bibnamefont {Brajard}},\
  and\ \bibinfo {author} {\bibfnamefont {P.}~\bibnamefont {Gallinari}},\
  }\bibfield  {title} {\bibinfo {title} {Learning dynamical systems from
  partial observations},\ }\href@noop {} {\bibfield  {journal} {\bibinfo
  {journal} {arXiv preprint arXiv:1902.11136}\ } (\bibinfo {year}
  {2019})}\BibitemShut {NoStop}%
\bibitem [{\citenamefont {Baydin}\ \emph {et~al.}(2018)\citenamefont {Baydin},
  \citenamefont {Barak}, \citenamefont {Radul},\ and\ \citenamefont
  {Siskind}}]{baydin2018autodiff}%
  \BibitemOpen
  \bibfield  {author} {\bibinfo {author} {\bibfnamefont {A.~G.}\ \bibnamefont
  {Baydin}}, \bibinfo {author} {\bibfnamefont {P.}~\bibnamefont {Barak}},
  \bibinfo {author} {\bibfnamefont {A.~A.}\ \bibnamefont {Radul}},\ and\
  \bibinfo {author} {\bibfnamefont {J.}~\bibnamefont {Siskind}},\ }\bibfield
  {title} {\bibinfo {title} {Automatic differentiation in machine learning: a
  survey},\ }\href@noop {} {\bibfield  {journal} {\bibinfo  {journal} {Journal
  of Machine Learning Research}\ }\textbf {\bibinfo {volume} {18}},\ \bibinfo
  {pages} {1–43} (\bibinfo {year} {2018})}\BibitemShut {NoStop}%
\bibitem [{\citenamefont {Gardiner}\ \emph {et~al.}(1985)\citenamefont
  {Gardiner} \emph {et~al.}}]{gardiner1985handbook}%
  \BibitemOpen
  \bibfield  {author} {\bibinfo {author} {\bibfnamefont {C.~W.}\ \bibnamefont
  {Gardiner}} \emph {et~al.},\ }\href@noop {} {\emph {\bibinfo {title}
  {Handbook of stochastic methods}}},\ Vol.~\bibinfo {volume} {3}\ (\bibinfo
  {publisher} {springer Berlin},\ \bibinfo {year} {1985})\BibitemShut {NoStop}%
\bibitem [{\citenamefont {Hochreiter}\ and\ \citenamefont
  {Schmidhuber}(1997)}]{hochreiter1997long}%
  \BibitemOpen
  \bibfield  {author} {\bibinfo {author} {\bibfnamefont {S.}~\bibnamefont
  {Hochreiter}}\ and\ \bibinfo {author} {\bibfnamefont {J.}~\bibnamefont
  {Schmidhuber}},\ }\bibfield  {title} {\bibinfo {title} {Long short-term
  memory},\ }\href@noop {} {\bibfield  {journal} {\bibinfo  {journal} {Neural
  computation}\ }\textbf {\bibinfo {volume} {9}},\ \bibinfo {pages} {1735}
  (\bibinfo {year} {1997})}\BibitemShut {NoStop}%
\bibitem [{\citenamefont {Sch{\"o}lkopf}(2019)}]{scholkopf2019causality}%
  \BibitemOpen
  \bibfield  {author} {\bibinfo {author} {\bibfnamefont {B.}~\bibnamefont
  {Sch{\"o}lkopf}},\ }\bibfield  {title} {\bibinfo {title} {Causality for
  machine learning},\ }\href@noop {} {\bibfield  {journal} {\bibinfo  {journal}
  {arXiv preprint arXiv:1911.10500}\ } (\bibinfo {year} {2019})}\BibitemShut
  {NoStop}%
\bibitem [{\citenamefont {Zaheer}\ \emph {et~al.}(2017)\citenamefont {Zaheer},
  \citenamefont {Kottur}, \citenamefont {Ravanbakhsh}, \citenamefont {Poczos},
  \citenamefont {Salakhutdinov},\ and\ \citenamefont {Smola}}]{zaheer2017deep}%
  \BibitemOpen
  \bibfield  {author} {\bibinfo {author} {\bibfnamefont {M.}~\bibnamefont
  {Zaheer}}, \bibinfo {author} {\bibfnamefont {S.}~\bibnamefont {Kottur}},
  \bibinfo {author} {\bibfnamefont {S.}~\bibnamefont {Ravanbakhsh}}, \bibinfo
  {author} {\bibfnamefont {B.}~\bibnamefont {Poczos}}, \bibinfo {author}
  {\bibfnamefont {R.~R.}\ \bibnamefont {Salakhutdinov}},\ and\ \bibinfo
  {author} {\bibfnamefont {A.~J.}\ \bibnamefont {Smola}},\ }\bibfield  {title}
  {\bibinfo {title} {Deep sets},\ }\href@noop {} {\bibfield  {journal}
  {\bibinfo  {journal} {Advances in neural information processing systems}\
  }\textbf {\bibinfo {volume} {30}} (\bibinfo {year} {2017})}\BibitemShut
  {NoStop}%
\bibitem [{\citenamefont {Xu}\ \emph {et~al.}(2018)\citenamefont {Xu},
  \citenamefont {Hu}, \citenamefont {Leskovec},\ and\ \citenamefont
  {Jegelka}}]{xu2018powerful}%
  \BibitemOpen
  \bibfield  {author} {\bibinfo {author} {\bibfnamefont {K.}~\bibnamefont
  {Xu}}, \bibinfo {author} {\bibfnamefont {W.}~\bibnamefont {Hu}}, \bibinfo
  {author} {\bibfnamefont {J.}~\bibnamefont {Leskovec}},\ and\ \bibinfo
  {author} {\bibfnamefont {S.}~\bibnamefont {Jegelka}},\ }\bibfield  {title}
  {\bibinfo {title} {How powerful are graph neural networks?},\ }\href@noop {}
  {\bibfield  {journal} {\bibinfo  {journal} {arXiv preprint arXiv:1810.00826}\
  } (\bibinfo {year} {2018})}\BibitemShut {NoStop}%
\bibitem [{\citenamefont {Jang}\ \emph {et~al.}(2017)\citenamefont {Jang},
  \citenamefont {Gu},\ and\ \citenamefont {Poole}}]{jang2016categorical}%
  \BibitemOpen
  \bibfield  {author} {\bibinfo {author} {\bibfnamefont {E.}~\bibnamefont
  {Jang}}, \bibinfo {author} {\bibfnamefont {S.}~\bibnamefont {Gu}},\ and\
  \bibinfo {author} {\bibfnamefont {B.}~\bibnamefont {Poole}},\ }\bibfield
  {title} {\bibinfo {title} {Categorical reparameterization with
  gumbel-softmax},\ }in\ \href@noop {} {\emph {\bibinfo {booktitle} {Fifth
  International Conference on Learning Representations, {ICLR}}}}\ (\bibinfo
  {publisher} {OpenReview.net},\ \bibinfo {year} {2017})\BibitemShut {NoStop}%
\bibitem [{\citenamefont {Lee}\ \emph {et~al.}(2018)\citenamefont {Lee},
  \citenamefont {Bahri}, \citenamefont {Novak}, \citenamefont {Schoenholz},
  \citenamefont {Pennington},\ and\ \citenamefont
  {Sohl{-}Dickstein}}]{lee2018deep}%
  \BibitemOpen
  \bibfield  {author} {\bibinfo {author} {\bibfnamefont {J.}~\bibnamefont
  {Lee}}, \bibinfo {author} {\bibfnamefont {Y.}~\bibnamefont {Bahri}}, \bibinfo
  {author} {\bibfnamefont {R.}~\bibnamefont {Novak}}, \bibinfo {author}
  {\bibfnamefont {S.~S.}\ \bibnamefont {Schoenholz}}, \bibinfo {author}
  {\bibfnamefont {J.}~\bibnamefont {Pennington}},\ and\ \bibinfo {author}
  {\bibfnamefont {J.}~\bibnamefont {Sohl{-}Dickstein}},\ }\bibfield  {title}
  {\bibinfo {title} {Deep neural networks as gaussian processes},\ }in\
  \href@noop {} {\emph {\bibinfo {booktitle} {Sixth International Conference on
  Learning Representations, {ICLR}}}}\ (\bibinfo  {publisher}
  {OpenReview.net},\ \bibinfo {year} {2018})\BibitemShut {NoStop}%
\bibitem [{\citenamefont {Garcia}\ \emph {et~al.}(2002)\citenamefont {Garcia},
  \citenamefont {Parravano}, \citenamefont {Cosenza}, \citenamefont
  {Jim{\'e}nez},\ and\ \citenamefont {Marcano}}]{garcia2002coupled}%
  \BibitemOpen
  \bibfield  {author} {\bibinfo {author} {\bibfnamefont {P.}~\bibnamefont
  {Garcia}}, \bibinfo {author} {\bibfnamefont {A.}~\bibnamefont {Parravano}},
  \bibinfo {author} {\bibfnamefont {M.}~\bibnamefont {Cosenza}}, \bibinfo
  {author} {\bibfnamefont {J.}~\bibnamefont {Jim{\'e}nez}},\ and\ \bibinfo
  {author} {\bibfnamefont {A.}~\bibnamefont {Marcano}},\ }\bibfield  {title}
  {\bibinfo {title} {Coupled map networks as communication schemes},\
  }\href@noop {} {\bibfield  {journal} {\bibinfo  {journal} {Physical Review
  E}\ }\textbf {\bibinfo {volume} {65}},\ \bibinfo {pages} {045201} (\bibinfo
  {year} {2002})}\BibitemShut {NoStop}%
\bibitem [{\citenamefont {Campbell}\ \emph {et~al.}(1954)\citenamefont
  {Campbell}, \citenamefont {Gurin},\ and\ \citenamefont
  {Miller}}]{campbell1954voter}%
  \BibitemOpen
  \bibfield  {author} {\bibinfo {author} {\bibfnamefont {A.}~\bibnamefont
  {Campbell}}, \bibinfo {author} {\bibfnamefont {G.}~\bibnamefont {Gurin}},\
  and\ \bibinfo {author} {\bibfnamefont {W.~E.}\ \bibnamefont {Miller}},\
  }\href@noop {} {\emph {\bibinfo {title} {The voter decides.}}}\ (\bibinfo
  {publisher} {Row, Peterson, and Co.},\ \bibinfo {year} {1954})\BibitemShut
  {NoStop}%
\bibitem [{\citenamefont {Karlebach}\ and\ \citenamefont
  {Shamir}(2008)}]{karlebach2008modelling}%
  \BibitemOpen
  \bibfield  {author} {\bibinfo {author} {\bibfnamefont {G.}~\bibnamefont
  {Karlebach}}\ and\ \bibinfo {author} {\bibfnamefont {R.}~\bibnamefont
  {Shamir}},\ }\bibfield  {title} {\bibinfo {title} {Modelling and analysis of
  gene regulatory networks},\ }\href@noop {} {\bibfield  {journal} {\bibinfo
  {journal} {Nature Reviews Molecular Cell Biology}\ }\textbf {\bibinfo
  {volume} {9}},\ \bibinfo {pages} {770} (\bibinfo {year} {2008})}\BibitemShut
  {NoStop}%
\bibitem [{\citenamefont {Brockmann}\ and\ \citenamefont
  {Helbing}(2013)}]{brockmann2013hidden}%
  \BibitemOpen
  \bibfield  {author} {\bibinfo {author} {\bibfnamefont {D.}~\bibnamefont
  {Brockmann}}\ and\ \bibinfo {author} {\bibfnamefont {D.}~\bibnamefont
  {Helbing}},\ }\bibfield  {title} {\bibinfo {title} {The hidden geometry of
  complex, network-driven contagion phenomena},\ }\href@noop {} {\bibfield
  {journal} {\bibinfo  {journal} {Science}\ }\textbf {\bibinfo {volume}
  {342}},\ \bibinfo {pages} {1337} (\bibinfo {year} {2013})}\BibitemShut
  {NoStop}%
\bibitem [{\citenamefont {Schaffter}\ \emph {et~al.}(2011)\citenamefont
  {Schaffter}, \citenamefont {Marbach},\ and\ \citenamefont
  {Floreano}}]{schaffter2011genenetweaver}%
  \BibitemOpen
  \bibfield  {author} {\bibinfo {author} {\bibfnamefont {T.}~\bibnamefont
  {Schaffter}}, \bibinfo {author} {\bibfnamefont {D.}~\bibnamefont {Marbach}},\
  and\ \bibinfo {author} {\bibfnamefont {D.}~\bibnamefont {Floreano}},\
  }\bibfield  {title} {\bibinfo {title} {Genenetweaver: in silico benchmark
  generation and performance profiling of network inference methods},\
  }\href@noop {} {\bibfield  {journal} {\bibinfo  {journal} {Bioinformatics}\
  }\textbf {\bibinfo {volume} {27}},\ \bibinfo {pages} {2263} (\bibinfo {year}
  {2011})}\BibitemShut {NoStop}%
\bibitem [{\citenamefont {Erd{\H{o}}s}\ and\ \citenamefont
  {R{\'e}nyi}(1960)}]{erdHos1960evolution}%
  \BibitemOpen
  \bibfield  {author} {\bibinfo {author} {\bibfnamefont {P.}~\bibnamefont
  {Erd{\H{o}}s}}\ and\ \bibinfo {author} {\bibfnamefont {A.}~\bibnamefont
  {R{\'e}nyi}},\ }\bibfield  {title} {\bibinfo {title} {On the evolution of
  random graphs},\ }\href@noop {} {\bibfield  {journal} {\bibinfo  {journal}
  {Publ. Math. Inst. Hung. Acad. Sci}\ }\textbf {\bibinfo {volume} {5}},\
  \bibinfo {pages} {17} (\bibinfo {year} {1960})}\BibitemShut {NoStop}%
\bibitem [{\citenamefont {Barab{\'a}si}\ and\ \citenamefont
  {Albert}(1999)}]{barabasi1999emergence}%
  \BibitemOpen
  \bibfield  {author} {\bibinfo {author} {\bibfnamefont {A.-L.}\ \bibnamefont
  {Barab{\'a}si}}\ and\ \bibinfo {author} {\bibfnamefont {R.}~\bibnamefont
  {Albert}},\ }\bibfield  {title} {\bibinfo {title} {Emergence of scaling in
  random networks},\ }\href@noop {} {\bibfield  {journal} {\bibinfo  {journal}
  {Science}\ }\textbf {\bibinfo {volume} {286}},\ \bibinfo {pages} {509}
  (\bibinfo {year} {1999})}\BibitemShut {NoStop}%
\bibitem [{\citenamefont {Rossi}\ and\ \citenamefont {Ahmed}(2015)}]{nr}%
  \BibitemOpen
  \bibfield  {author} {\bibinfo {author} {\bibfnamefont {R.}~\bibnamefont
  {Rossi}}\ and\ \bibinfo {author} {\bibfnamefont {N.}~\bibnamefont {Ahmed}},\
  }\bibfield  {title} {\bibinfo {title} {The network data repository with
  interactive graph analytics and visualization},\ }in\ \href@noop {} {\emph
  {\bibinfo {booktitle} {Proceedings of the AAAI Conference on Artificial
  Intelligence}}},\ Vol.~\bibinfo {volume} {29}\ (\bibinfo  {publisher} {{AAAI}
  Press},\ \bibinfo {year} {2015})\ pp.\ \bibinfo {pages}
  {4292--4293}\BibitemShut {NoStop}%
\bibitem [{\citenamefont {Zhang}\ \emph {et~al.}(2020)\citenamefont {Zhang},
  \citenamefont {Dong}, \citenamefont {Zhang}, \citenamefont {Chen},
  \citenamefont {Yao},\ and\ \citenamefont {Han}}]{zhang2020investigating}%
  \BibitemOpen
  \bibfield  {author} {\bibinfo {author} {\bibfnamefont {J.}~\bibnamefont
  {Zhang}}, \bibinfo {author} {\bibfnamefont {L.}~\bibnamefont {Dong}},
  \bibinfo {author} {\bibfnamefont {Y.}~\bibnamefont {Zhang}}, \bibinfo
  {author} {\bibfnamefont {X.}~\bibnamefont {Chen}}, \bibinfo {author}
  {\bibfnamefont {G.}~\bibnamefont {Yao}},\ and\ \bibinfo {author}
  {\bibfnamefont {Z.}~\bibnamefont {Han}},\ }\bibfield  {title} {\bibinfo
  {title} {Investigating time, strength, and duration of measures in
  controlling the spread of covid-19 using a networked meta-population model},\
  }\href@noop {} {\bibfield  {journal} {\bibinfo  {journal} {Nonlinear
  Dynamics}\ }\textbf {\bibinfo {volume} {101}},\ \bibinfo {pages} {1789}
  (\bibinfo {year} {2020})}\BibitemShut {NoStop}%
\bibitem [{\citenamefont {Chen}\ \emph {et~al.}(2020)\citenamefont {Chen},
  \citenamefont {Zhang}, \citenamefont {Zhang}, \citenamefont {Du},
  \citenamefont {Hu}, \citenamefont {Wang},\ and\ \citenamefont
  {Zhu}}]{chen2020network}%
  \BibitemOpen
  \bibfield  {author} {\bibinfo {author} {\bibfnamefont {M.}~\bibnamefont
  {Chen}}, \bibinfo {author} {\bibfnamefont {J.}~\bibnamefont {Zhang}},
  \bibinfo {author} {\bibfnamefont {Z.}~\bibnamefont {Zhang}}, \bibinfo
  {author} {\bibfnamefont {L.}~\bibnamefont {Du}}, \bibinfo {author}
  {\bibfnamefont {Q.}~\bibnamefont {Hu}}, \bibinfo {author} {\bibfnamefont
  {S.}~\bibnamefont {Wang}},\ and\ \bibinfo {author} {\bibfnamefont
  {J.}~\bibnamefont {Zhu}},\ }\bibfield  {title} {\bibinfo {title} {Inference
  for network structure and dynamics from time series data via graph neural
  network},\ }\href@noop {} {\bibfield  {journal} {\bibinfo  {journal} {arxiv
  preprint: arXiv:2001.06576}\ } (\bibinfo {year} {2020})}\BibitemShut
  {NoStop}%
\bibitem [{\citenamefont {Pearl}\ and\ \citenamefont
  {Mackenzie}(2018)}]{pearl2018book}%
  \BibitemOpen
  \bibfield  {author} {\bibinfo {author} {\bibfnamefont {J.}~\bibnamefont
  {Pearl}}\ and\ \bibinfo {author} {\bibfnamefont {D.}~\bibnamefont
  {Mackenzie}},\ }\href@noop {} {\emph {\bibinfo {title} {The book of why: the
  new science of cause and effect}}}\ (\bibinfo  {publisher} {Basic Books},\
  \bibinfo {year} {2018})\BibitemShut {NoStop}%
\end{thebibliography}%

\end{document}